\documentclass[twocolumn, tighten, times]{aastex631}

\shorttitle{MIRI Imaging and Coronagraphic Flux Calibration}
\shortauthors{Gordon et al.}

\begin{document}

\title{The James Webb Space Telescope Absolute Flux Calibration. II. Mid-Infrared Instrument Imaging and Coronagraphy}

\author[0000-0001-5340-6774]{Karl D.\ Gordon}
\affiliation{Space Telescope Science Institute, 3700 San Martin
  Drive, Baltimore, MD, 21218, USA}
\affiliation{Sterrenkundig Observatorium, Universiteit Gent,
  Gent, Belgium}

\author[0000-0003-4520-1044]{G.~C.\ Sloan}
\affiliation{Space Telescope Science Institute, 3700 San Martin
  Drive, Baltimore, MD, 21218, USA}
  \affiliation{Department of Physics and Astronomy, University
  of North Carolina, Chapel Hill, NC 27599-3255, USA}

\author[0000-0003-4801-0489]{Macarena Garcia Marin}
\affiliation{European Space Agency (ESA), ESA Office, Space Telescope Science Institute, 3700 San Martin Drive, Baltimore, MD 21218, USA}

\author[0000-0001-9673-7397]{Mattia Libralato}
\affil{AURA for the European Space Agency (ESA), Space Telescope Science Institute, 3700 San Martin Drive, Baltimore, MD 21218, USA}
\affil{INAF, Osservatorio Astronomico di Padova, Vicolo dell’Osservatorio 5, Padova, I-35122, Italy}

\author[0000-0003-2303-6519]{George Rieke}
\affiliation{Steward Observatory, University of Arizona, Tucson,
  AZ 85721, USA}

\author[0000-0003-3184-0873]{Jonathan A.\ Aguilar}
\affiliation{Space Telescope Science Institute, 3700 San Martin Drive, Baltimore, MD, 21218, USA}

\author[0000-0001-9806-0551]{Ralph Bohlin}
\affiliation{Space Telescope Science Institute, 3700 San Martin
  Drive, Baltimore, MD, 21218, USA}

\author[0000-0002-7698-3002]{Misty Cracraft}
\affiliation{Space Telescope Science Institute, 3700 San Martin
  Drive, Baltimore, MD, 21218, USA}

\author[0000-0001-9462-5543]{Marjorie Decleir}
\altaffiliation{ESA Research Fellow}
\affiliation{European Space Agency (ESA), ESA Office, Space Telescope Science Institute, 3700 San Martin Drive, Baltimore, MD 21218, USA}

\author[0000-0001-8612-3236]{Andras Gaspar}
\affiliation{Steward Observatory, University of Arizona, Tucson,
  AZ 85721, USA}

\author[0000-0002-7612-0469]{Sarah Kendrew}
\affiliation{European Space Agency (ESA), ESA Office, Space Telescope Science Institute, 3700 San Martin Drive, Baltimore, MD 21218, USA}

\author[0000-0002-9402-186X]{David R.\ Law}
\affiliation{Space Telescope Science Institute, 3700 San Martin Drive, Baltimore, MD, 21218, USA}

\author[0000-0002-6296-8960]{Alberto Noriega-Crespo}
\affiliation{Space Telescope Science Institute, 3700 San Martin Drive, Baltimore, MD, 21218, USA}

\author[0000-0001-9367-0705]{Michael Regan}
\affiliation{Space Telescope Science Institute, 3700 San Martin
  Drive, Baltimore, MD, 21218, USA}


\begin{abstract}
The absolute flux calibration of the Mid-Infrared Instrument Imaging and Coronagraphy is based on observations of multiple stars taken during the first 2.5 years of JWST operations. 
The observations were designed to ensure that the flux calibration is valid for a range of flux densities, different subarrays, and different types of stars.
The flux calibration was measured by combining observed aperture photometry corrected to infinite aperture with predictions based on previous observations and models of stellar atmospheres.
A subset of these observations were combined with model point-spread-functions to measure the corrections to infinite aperture.
Variations in the calibration factor with time, flux density, background level, type of star, subarray, integration time, rate, and well depth were investigated, and the only significant variations were with time and subarray.
Observations of the same star taken approximately every month revealed a modest time-dependent response loss seen mainly at the longest wavelengths.
This loss is well characterized by a decaying exponential with a time constant of $\sim$200 days. 
After correcting for the response loss, the band-dependent scatter around the corrected average (aka repeatability) was found to range from 0.1 to 1.2\%.
Signals in observations taken with different subarrays can be lower by up to 3.4\% compared to FULL frame.
After correcting for the time and subarray dependencies, the scatter in the calibration factors measured for individual stars ranges from 1 to 4\% depending on the band.
The formal uncertainties on the flux calibration averaged for all observations are 0.3 to 1.0\%, with longer-wavelength bands generally having larger uncertainties.
\end{abstract}


\section{Introduction} \label{sec:intro}

Absolute calibration of astronomical observations is essential to many astrophysical investigations.
A holistic approach is taken to calibrate all the James Webb Space Telescope (JWST) \citep[JWST, ][]{Gardner23} instruments consistently.
At the same time, accurate calibration requires extending infrared calibration sequences to much fainter levels than has been the case previously before JWST \citep{Gordon22}.
This paper describes the steps to calibrate the Mid-Infrared Instrument \citep[MIRI,][]{Rieke15, Gillian23}; it is one of a series of papers covering all the JWST instruments.

MIRI on the James Webb Space Telescope \citep[JWST,][]{Gardner23} provides imaging, coronagraphic, and spectroscopic observations in the mid-infrared from 5 to 28.8~\micron.
The absolute flux calibration of MIRI is based on observations of stars with well-modeled spectral energy distributions that have been taken as part of the overall JWST absolute flux-calibration program \citep{Gordon22}.
As for all the JWST instruments, the MIRI observations target stars of three different types and include multiple stars of each type.
This coverage allows for both random and systematic uncertainties to be quantified.
The combination of the cycle 1 and 2 JWST absolute flux-calibration program was constructed to enable accuracies of at least 5\% and 10\% for MIRI imaging and spectroscopy, respectively.
Providing even higher accuracies is a goal of the program to enhance the JWST science.
An explicit goal is to support science investigations using point and extended sources.
See \citet{Gordon22} for the details of the integrated JWST absolute flux calibration program for all JWST instruments.

\begin{figure*}[tbp]
\epsscale{1.1}
\plotone{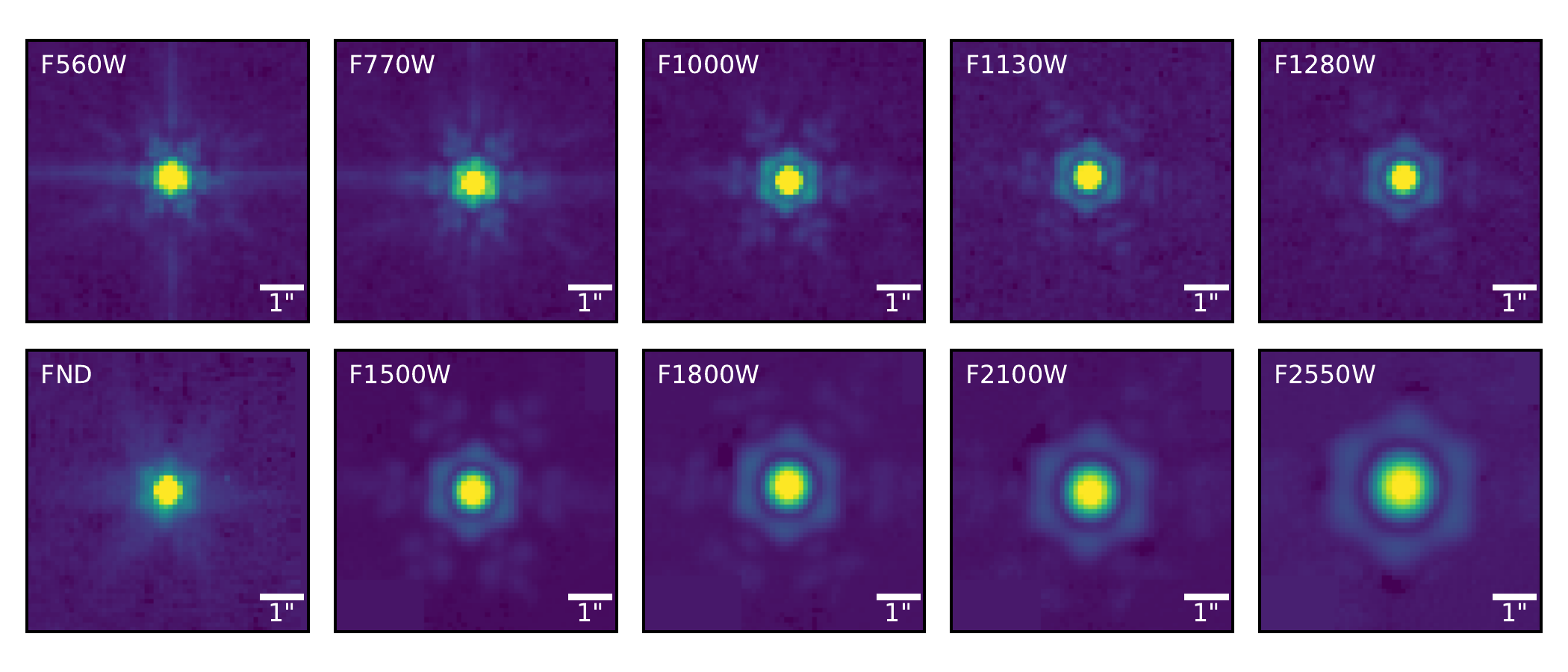} \\
\epsscale{0.55}
\plotone{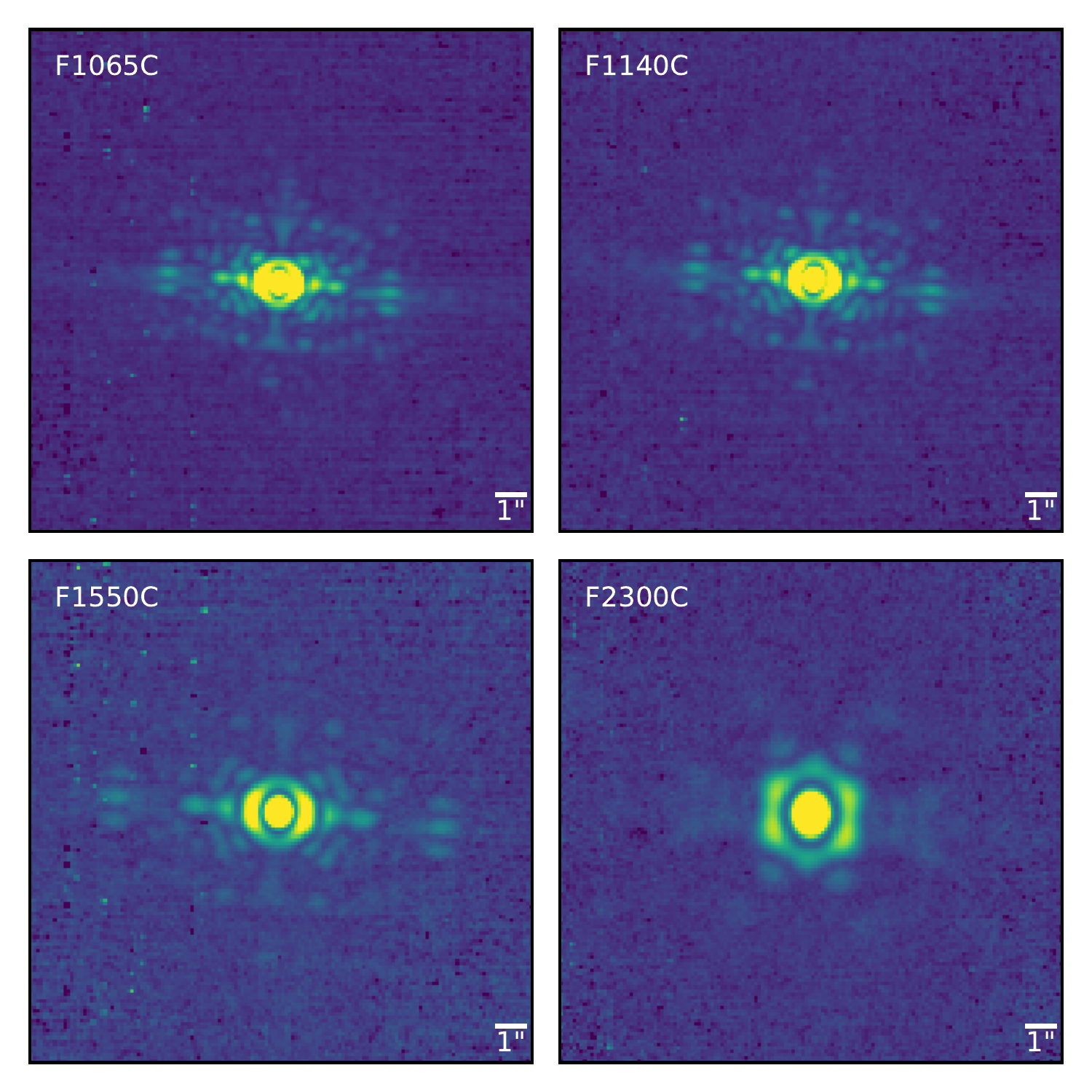}
\caption{Example images for the imaging and coronagraphic filters to illustrate the data quality and general characteristics of MIRI point source observations. 
The targets include BD+60~1753 for the imaging bands from F560W to F1280W, $\delta$~UMi for the rest of the imaging bands, and HD~2811 for the coronagraphic bands.  
The image in the neutral density filter (FND) is smoother compared to the other imaging bands due to its wide bandpass.
The coronagraphic observations were taken well away from the coronagraphic centers, and so these are the unocculted PSFs.
\label{fig:eximages}}
\end{figure*}

\begin{deluxetable*}{lrll}[tbp]
\tablewidth{0pt}
\tabletypesize{\scriptsize}
\tablecaption{Hot Star Observations\label{tab:hotstarobs}}
\tablehead{\colhead{Name} & \colhead{PID} & \colhead{Subarray} & \colhead{Bands}}
\startdata
10 Lac & 1524 & FULL & FND \\
 & 4497 & SUB64 & F1000W, F1130W, F1280W, F1500W, F1800W, F2100W, F2550W \\
G 191-B2B & 1537 & FULL & F560W, F770W, F1000W, F1130W, F1280W, F1500W \\
GD 153 & 1537 & FULL & F560W, F770W, F1000W \\
 & 4497 & FULL & F560W, F770W, F1000W \\
GD 71 & 1537 & FULL & F560W, F770W, F1000W \\
LAWD 87 & 4497 & FULL & F560W, F770W, F1000W \\
WD 1057+719 & 4497 & FULL & F560W, F770W \\
$\lambda$ Lep & 4497 & SUB64 & F1130W, F1280W, F1500W, F1800W, F2100W, F2550W \\
$\mu$ Col & 4497 & FULL & FND \\
 &  & SUB64 & F1000W, F1130W, F1280W, F1500W, F1800W, F2100W, F2550W \\
\enddata
\end{deluxetable*}

\begin{deluxetable*}{lrll}[tbp]
\tablewidth{0pt}
\tabletypesize{\scriptsize}
\tablecaption{Solar Analog Observations\label{tab:solaranalogsobs}}
\tablehead{\colhead{Name} & \colhead{PID} & \colhead{Subarray} & \colhead{Bands}}
\startdata
16 Cyg B & 1538 & FULL & FND \\
 &  & SUB64 & F1130W, F1280W, F1500W, F1800W, F2100W, F2550W \\
C26202 & 4498 & FULL & F560W, F770W, F1000W \\
GSPC P177-D & 1538 & BRIGHTSKY & F560W, F770W, F1000W, F1130W, F1280W, F1500W, F1800W \\
GSPC P330-E & 1538 & BRIGHTSKY & F560W, F770W, F1000W, F1130W, F1280W, F1500W, F1800W \\
 & 4498 & FULL & F560W, F770W, F1000W, F1130W, F1280W, F1500W, F1800W \\
 &  & MASK1065 & F1065C \\
 &  & MASK1140 & F1140C \\
HD 106252 & 1538 & SUB64 & F1000W, F1130W, F1280W, F1500W, F1800W, F2100W, F2550W \\
HD 1452331 & 4498 & SUB64 & F1000W, F1130W, F1280W, F1500W, F1800W, F2100W, F2550W \\
HD 167060 & 1538 & FULL & FND \\
 &  & MASK1065 & F1065C \\
 &  & MASK1140 & F1140C \\
 &  & MASK1550 & F1550C \\
 &  & MASKLYOT & F2300C \\
 &  & SUB64 & F770W, F1000W, F1130W, F1280W, F1500W, F1800W, F2100W, F2550W \\
HD 37962 & 1538 & FULL & FND \\
 &  & SUB64 & F1000W, F1130W, F1280W, F1500W, F1800W, F2100W, F2550W \\
 & 4578 & SUB64 & F1280W, F1500W, F1800W, F2100W, F2550W \\
HR 6538 & 4498 & FULL & FND \\
 &  & SUB64 & F1130W, F1280W, F1500W, F1800W, F2100W, F2550W \\
SNAP-2 & 4498 & FULL & F560W, F770W, F1000W \\
\enddata
\end{deluxetable*}

This paper presents the analysis of MIRI imaging and coronagraphic data taken over the first 2.5 years of JWST operations (i.e., Commissioning, cycle 1, and cycle 2).
The overall goal of this analysis is to provide the flux calibration that converts the measured DN~s$^{-1}$~pixel$^{-1}$ values to physical MJy~sr$^{-1}$ surface brightness units for all the imaging and coronagraphic filters.
The sources for which flux densities can be predicted at the accuracies required are generally limited to stars which are straightforward to model.
A number of hot stars, A dwarfs, and Solar analogs with a range of flux densities were observed with all the MIRI imaging and coronagraphic filters.
These observations were taken with different subarrays, different integration times, and were spread throughout the observing period.
A critical part of this analysis is the derivation of aperture corrections to infinite apertures because the stellar model predictions are for the total flux  densities.
In addition, this correction explicitly allows the flux calibration to support both point and extended source science simultaneously by calibrating images to surface brightness units \citep{Gordon22}.

\begin{deluxetable*}{lrll}[tbp]
\tablewidth{0pt}
\tabletypesize{\scriptsize}
\tablecaption{A Dwarf Observations\label{tab:adwarfobs}}
\tablehead{\colhead{Name} & \colhead{PID} & \colhead{Subarray} & \colhead{Bands}}
\startdata
2MASS J17430448+6655015 & 1027 & FULL & F560W, F770W, F1000W, F1130W, F1280W, F1500W \\
 & 1536 & FULL & F560W, F770W, F1000W, F1130W, F1280W, F1500W \\
2MASS J17571324+6703409 & 1533 & BRIGHTSKY & F770W \\
 &  & FULL & F770W \\
 &  & SUB128 & F770W \\
 &  & SUB256 & F770W \\
 &  & SUB64 & F770W \\
 & 1536 & BRIGHTSKY & F560W \\
 & 4496 & BRIGHTSKY & F560W, F770W, F1000W, F1130W, F1280W, F1500W, F1800W, F2100W \\
2MASS J18022716+6043356 & 1536 & BRIGHTSKY & F560W, F770W, F1000W, F1130W, F1280W, F1500W, F1800W \\
 & 4488 & BRIGHTSKY & F1280W \\
 &  & FULL & F1280W \\
 &  & SUB128 & F1280W \\
 &  & SUB256 & F1280W \\
 &  & SUB64 & F1280W \\
BD+60 1753 & 1027 & FULL & F1500W, F1800W, F2100W, F2550W \\
 &  & SUB256 & F560W, F770W, F1000W, F1500W \\
 & 1045 & MASK1065 & F1065C \\
 &  & MASK1140 & F1140C \\
 &  & MASK1550 & F1550C \\
 &  & MASKLYOT & F2300C \\
 & 1536 & SUB256 & F560W, F770W, F1000W, F1130W, F1280W, F1500W, F1800W \\
 & 1539 & SUB256 & F770W \\
 & 4499 & SUB256 & F560W, F770W, F1000W, F1130W, F1280W, F1500W, F1800W, F2100W, F2550W \\
 & 6607 & SUB256 & F560W, F770W, F1000W, F1130W, F1280W, F1500W, F1800W, F2100W, F2550W \\
HD 101452 & 4496 & BRIGHTSKY & F1500W, F1800W, F2100W, F2550W \\
HD 163466 & 1027 & SUB64 & F1000W, F1130W, F1280W, F1500W \\
 & 1536 & FULL & FND \\
 &  & SUB64 & F1000W, F1130W, F1280W, F1500W, F1800W, F2100W, F2550W \\
 & 1539 & FULL & FND \\
 & 6607 & FULL & FND \\
HD 180609 & 1536 & SUB128 & F560W, F770W, F1000W, F1130W, F1280W, F1500W, F1800W, F2100W, F2550W \\
HD 2811 & 1523 & SUB64 & F770W, F1000W, F1130W, F1280W, F1500W, F1800W, F2100W, F2550W \\
 & 1536 & FULL & FND \\
 &  & MASK1065 & F1065C \\
 &  & MASK1140 & F1140C \\
 &  & MASK1550 & F1550C \\
 &  & MASKLYOT & F2300C \\
 &  & SUB64 & F770W, F1000W, F1130W, F1280W, F1500W, F1800W, F2100W, F2550W \\
 & 4496 & BRIGHTSKY & F1500W, F1800W, F2100W, F2550W \\
 &  & FULL & FND \\
 &  & MASK1550 & F1550C \\
 &  & MASKLYOT & F2300C \\
HD 55677 & 4496 & SUB256 & F560W, F770W, F1000W, F1130W, F1280W, F1500W, F1800W, F2100W, F2550W \\
HR 5467 & 4496 & FULL & FND \\
 &  & SUB256 & F1500W, F1800W, F2100W, F2550W \\
del UMi & 1524 & FULL & FND \\
 & 1536 & FULL & FND \\
 &  & MASK1065 & F1065C \\
 &  & MASK1140 & F1140C \\
 &  & MASK1550 & F1550C \\
 &  & MASKLYOT & F2300C \\
 &  & SUB64 & F1130W, F1500W, F1800W, F2100W, F2550W \\
 & 4578 & SUB64 & F1500W, F1800W, F2100W, F2550W \\
\enddata
\end{deluxetable*}

\S\ref{sec:data} provides the details of the observations, the derivation of the apertures and aperture corrections, how the aperture photometry was performed, the flux densities predicted from models, and the zero-magnitude flux densities in the Sirius-Vega system.
The calculation of the calibration factors is given in \S\ref{sec:results} along with the measurement of the temporal and subarray dependence of these factors.
In addition, \S\ref{sec:results} illustrates the lack of any significant dependencies on flux density, background, type of star, central-pixel rate, central-pixel well depth, and integration time.
\S\ref{sec:summary} summarizes the results, and the appendix provides plots of the calibration factors from each star plotted versus all the dependencies investigated.

\section{Data \label{sec:data}}

The main observations used for this work are those taken specifically for the absolute flux calibration in cycle 1 and 2 (PIDs: 1523, 1524, 1536, 1537, 1538, 1539, 4488, 4496, 4497, 4498, 4499, and 4578).
A small amount of data taken in cycle 3 that was available at the time of this work are included (PID: 6607).
These data were supplemented with observations taken during Commissioning that provide the preliminary absolute flux calibration (PID: 1027 and 1045).
All of the targets were from the overall JWST absolute flux-calibration program and are highly vetted hot stars, A dwarfs, and Solar analog stars \citep{Gordon22}.
The exposure times were chosen to obtain data with signal-to-noise (S/N) ratios of 200 or better.
Tables~\ref{tab:hotstarobs}, \ref{tab:solaranalogsobs}, and \ref{tab:adwarfobs}, give the names of the stars, program identifications (PIDs), subarrays, and filters for the observations.
A small number of observations from these programs suffered from various issues (e.g., failed guide-star acquisition, lost tracking, etc.), and these observations were not used in this work, nor are they listed in the tables.

The data were reduced using the JWST pipeline version 1.15.1 with the reference files specified by calibration reference data system pmap 1256, specifically using the CALWEBB\_DETECTOR1, CALWEBB\_IMAGE2, and CALWEBB\_IMAGE3 stages.
The standard steps for imaging data were performed except as described below.
For the CALWEBB\_DETECTOR1, the electromagnetic interference correction (emicorr) was not applied, and for the jump detection step (jump), the after-jump flagging and shower detection were skipped.
For CALWEBB\_IMAGE3, the outlier detection (outlier\_detection), refinement of the relative astrometry (tweakreg), background matching (skymatch), and source catalog generation (source\_catalog) steps were also skipped.
In addition, the image combination step (resample) was run with a square kernel and the exposure time (exptime) weighting.
These modifications to the standard reductions were made as  well-exposed, high S/N observations with standard dithers do not benefit from some of the steps.

For the coronagraphic data, the data reduction was the same as imaging including the final mosaicking (e.g., CALWEBB\_IMAGE3 instead of CALWEBB\_CORON3).
The motivation for using standard image mosaicking is that these observations were taken with the source placed well away from the center of the coronagraph and thus resemble imaging observations more than typical coronagraphic observations.
To obtain reasonable parameters for the CALWEBB\_IMAGE3 pipeline steps for coronagraphic data, the filter in the calibration exposures (``cal'' files) was changed to the closest-in-wavelength imaging filter.

In addition to the standard processing to produce mosaics, we also generated mosaics with custom image-based background subtraction for each exposure before mosaicking.
The custom background for each band was created by averaging the stack of all the exposures in an observation aligned in detector coordinates.
Before obtaining this mean, the region within the full width at half maximum (FWHM) of the source was masked to ensure the mean background did not include any of the source signal.
These background-subtracted images and mosaics are important in cases where the background is strong and/or has significant structure, which happens in all of the coronagraphic bands due to significant scattered light \citep{Boccaletti22}.
The F2550W imaging band is affected similarly.

Fig.~\ref{fig:eximages} shows example mosaics created using the custom image-based background subtraction.

\subsection{Apertures and Corrections}
\label{sec:apcor}

\begin{figure*}[tbp]
\epsscale{1.1}
\plottwo{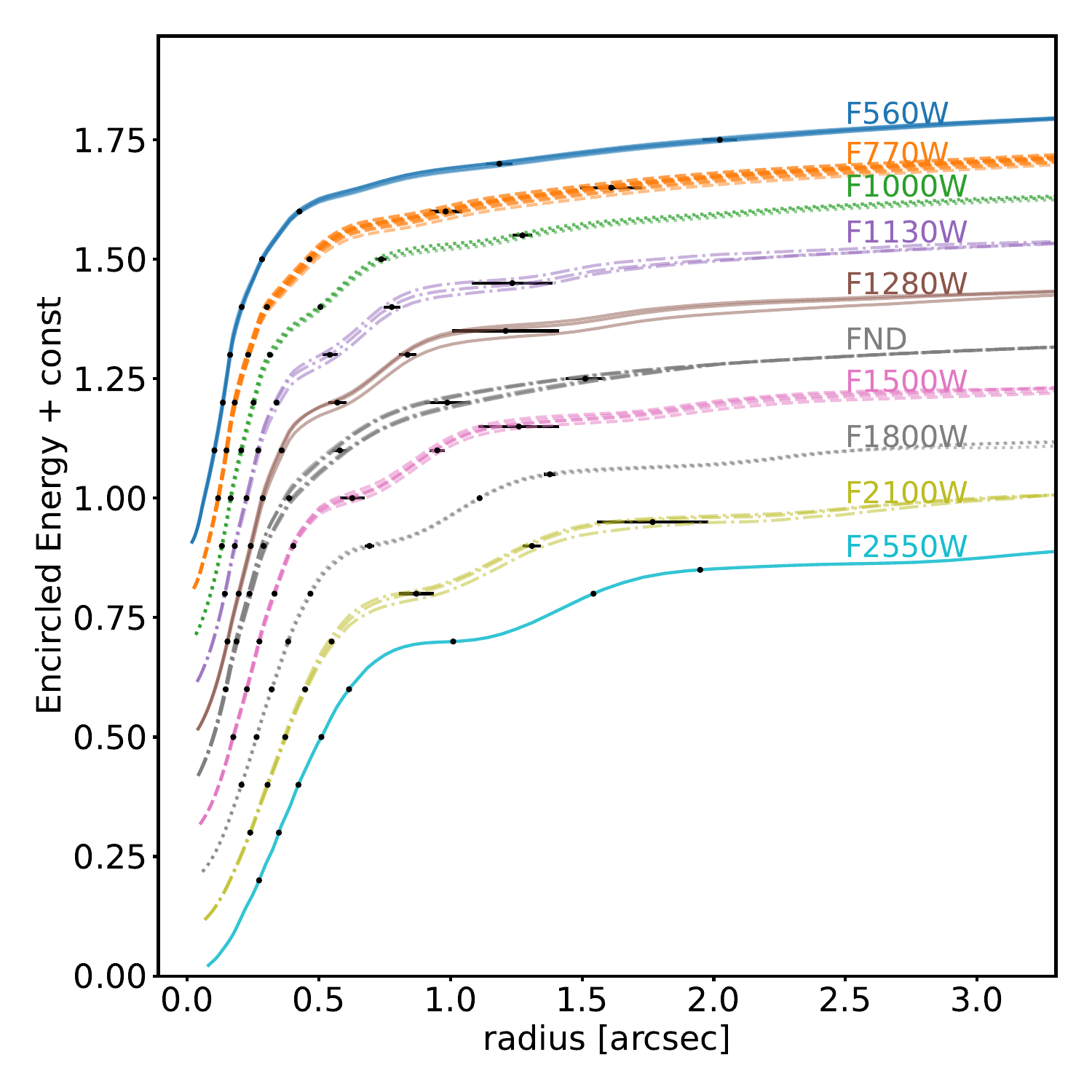}{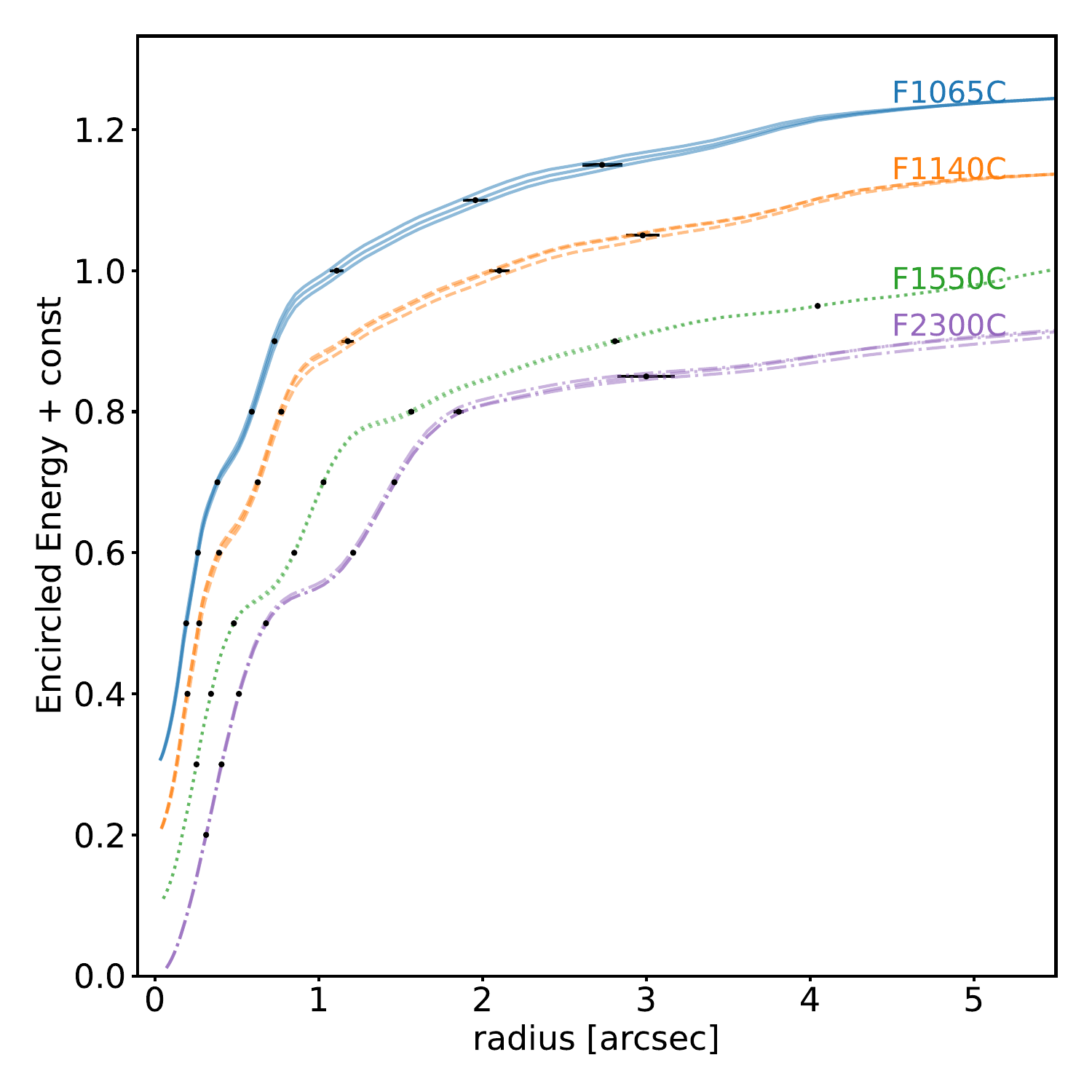}
\caption{The radial encircled energies for the imaging (left) and coronagraphic (right) bands.
Subsequent bands are offset by a factor of 0.1.
Multiple radial encircled energies are plotted for each band using repeated observations except F2550W, which is based on the WebbPSF model at all radii. 
The black points give the radii for 0.2, 0.3, 0.4, 0.5, 0.6, 0.7, 0.8, and 0.85 encircled energies with their uncertainties as horizontal lines.
The radii uncertainties are based on the scatter of the radii computed from independent observations.
The coronagraphic profiles are broader than the imaging profiles, and the Lyot F2300C profile is narrower than expected given an extrapolation of the behavior of the three 4QPM coronagraphic profiles (F1065C, F1140C, and F1550C).
This is due to the difference in type of coronagraph.
\label{fig:ee_example}}
\end{figure*}

The photometric apertures and background annuli for measuring the flux of each star were set based on the point-spread-functions (PSFs) in each band.
Specifically, radial profiles of encircled energy were used, with the aperture radius ($r_a$) set to include 70\% of the flux and the background annulus set between the radii that include 80\% and 85\% of the flux ($r_{b1}$ and $r_{b2}$).
The background annulus includes contributions from the actual background and from the source itself due to the PSF.
Thus, the aperture corrections necessarily depend on the background annulus parameters in addition to the object aperture radius.
With these radii values, the aperture correction to infinite aperture can be computed as
\begin{equation}
A_\mathrm{cor} = \frac{1}{0.7 - S_\mathrm{bkg} \pi r_a^2}
\end{equation}
where the background surface brightness per pixel due to the PSF wings is
\begin{equation}
S_\mathrm{bkg} = \frac{0.85 - 0.8}{\pi (r_{b2}^2 - r_{b1}^2)} .
\end{equation}

The radial encircled energy profiles were computed using a combination of observed and model PSFs.
It is not possible to create PSFs from observations alone as determining the background level far from the source is challenging given variations in the background due to residual detector and instrument effects, other astronomical sources, and measurement noise.
Using a model for the PSF based on the optics of the telescope and instrument is possible, but the structure at small radii in the observations does not match the model well, especially at the shorter MIRI wavelengths. 
One known detector issue that may affect the core is the ``brighter-fatter effect'' \citep[BFE;][]{Argyriou23}.
This effect arises when large differences in accumulated charge between neighboring pixels results in the diffusion of some charge to pixels with lower accumulated charge.
\S\ref{sec:detector} investigates the impact of BFE on the aperture photometry.
The model PSF was created using WebbPSF v1.3.0 \citep{Perrin12, Perrin14} using the option that includes detector artifacts including the cruciform \citep{Gaspar21, Dicken24}.

Hence, the MIRI PSFs are generated by using the observed radial PSFs inside a given radius and the model PSF outside this radius.
We did not use the existing effective PSFs (ePSFs) because they were not designed to measure the encircled energy due to smoothing and their normalization to unity area at a finite radius \citep{Libralato24}.
For the imaging filters, observations of BD+60~1753 were used for the observed PSF as it had multiple, high S/N observations with no visually obvious sources within 10\arcsec.
For the FND filter, $\delta$~UMi was used.
For the coronagraphic filters, observations of HD 2811, $\delta$~UMi, and 16 Cyg B were used.
The observed radial PSFs are multiplicatively corrected to match the model PSF at a given radius.
The radius where the PSFs transition between the observed and the model PSFs was determined based on visually inspecting the observed radial profiles and choosing the largest radius at each wavelength that did not show obvious extraneous structure (i.e., encircled energy decreasing with increasing radius).
The radii used for all bands were approximately 40 pixels, which corresponds to $\sim$4.5$\arcsec$.
It was not possible to determine a high-quality observed PSF for F2550W due to large variations in the background.
Hence, the model PSF is adopted for all radii for this filter.
This solution for F2550W is reasonable given that the difference between the model PSFs and observed PSFs decreases with wavelength and becomes quite small for F1800W and F2100W.  
Fig.~\ref{fig:ee_example} shows how the final encircled energies depend on wavelength for all of the bands.
The increasing PSF width with wavelength is clearly seen along with the signature of the first and second Airy rings.
The FND band shows smoother variation than other imaging bands because its wide bandpass washes out the Airy rings.
The coronagraphic filters have larger widths than imaging bands with similar wavelengths due to the extra coronagraphic optical elements \citep{Perrin18,Boccaletti15}.
The uncertainties are determined from the multiple observations in the same filter.
The radii are larger than expected for the F560W and F770W filters compared to the behavior of the F1000W and longer filters due to the extra cruciform component that deflects light below 10~\micron\ inside the detector material to larger radii.

\begin{deluxetable}{lcccc}[tbp]
\tablewidth{0pt}
\tabletypesize{\scriptsize}
\tablecaption{Apertures and Corrections\label{tab:apcor}}
\tablehead{\colhead{Band} & \colhead{$r_a$} & \colhead{$r_{b1}$} & \colhead{$r_{b2}$} & \colhead{$A_\mathrm{cor}$} \\
 & \colhead{(pix)} & \colhead{(pix)} & \colhead{(pix)} & }
\startdata
F560W & $3.87 \pm 0.06$ & $10.78 \pm 0.44$ & $18.39 \pm 0.59$ & $1.436 \pm 0.0004$ \\
F770W & $4.22 \pm 0.10$ & $8.92 \pm 0.57$ & $14.64 \pm 1.07$ & $1.442 \pm 0.0014$ \\
F1000W & $4.60 \pm 0.08$ & $6.70 \pm 0.19$ & $11.58 \pm 0.34$ & $1.453 \pm 0.0007$ \\
F1130W & $4.92 \pm 0.26$ & $7.06 \pm 0.29$ & $11.22 \pm 1.40$ & $1.465 \pm 0.0124$ \\
F1280W & $5.18 \pm 0.32$ & $7.61 \pm 0.30$ & $10.99 \pm 1.84$ & $1.483 \pm 0.0206$ \\
FND & $5.27 \pm 0.26$ & $8.98 \pm 0.78$ & $13.74 \pm 0.69$ & $1.455 \pm 0.0015$ \\
F1500W & $5.69 \pm 0.43$ & $8.63 \pm 0.26$ & $11.45 \pm 1.39$ & $1.497 \pm 0.0190$ \\
F1800W & $6.29 \pm 0.17$ & $10.09 \pm 0.04$ & $12.52 \pm 0.20$ & $1.506 \pm 0.0021$ \\
F2100W & $7.91 \pm 0.61$ & $11.90 \pm 0.31$ & $16.07 \pm 1.93$ & $1.492 \pm 0.0171$ \\
F2550W & $9.18 \pm 0.00$ & $14.03 \pm 0.00$ & $17.71 \pm 0.00$ & $1.506 \pm 0.0000$ \\
F1065C & $10.07 \pm 0.39$ & $17.77 \pm 0.68$ & $24.80 \pm 1.11$ & $1.464 \pm 0.0010$ \\
F1140C & $10.68 \pm 0.35$ & $19.10 \pm 0.58$ & $27.06 \pm 0.92$ & $1.461 \pm 0.0006$ \\
F1550C & $14.21 \pm 0.16$ & $25.52 \pm 0.25$ & $36.77 \pm 0.05$ & $1.459 \pm 0.0014$ \\
F2300C & $13.27 \pm 0.08$ & $16.84 \pm 0.29$ & $27.25 \pm 1.59$ & $1.470 \pm 0.0066$ \\
\enddata
\tablenotetext{\star}{F2550W does not have measured uncertainties as the model PSF was used for all radii.}
\end{deluxetable}

To determine the absolute flux calibration, we use radii for 70\% of the encircled energy with the sky annulus between the radii for 80 and 85\% of the encircled energies.  
Table~\ref{tab:apcor} gives the radii and aperture corrections for these choices for all of the bands.
The uncertainties in radii are strongly influenced by the steepness of the encircled energy profile.
For example, the $r_{b2}$ uncertainties are often significantly larger than the $r_{b1}$ as the 85\% points on the encircled energy curves are much less steep than the 80\% points.
The reference file for the aperture correction in the JWST pipeline provides the aperture corrections for encircled energies between 0.1 and 0.8 using background annuli defined by 80 and 85~\% of the encircled energies\footnote{\url{https://jwst-crds.stsci.edu/}}.

\begin{rotatetable*}
\begin{deluxetable*}{ccccccccccccccc}
\tablecaption{Photometry}
\label{tab:phot}
\tablefontsize{\scriptsize}
\tablehead{\colhead{name} & \colhead{PID} & \colhead{srctype} & \colhead{filter} & \colhead{subarray} & \colhead{time} & \colhead{pixrate} & \colhead{pixwelldepth} & \colhead{inttime} & \colhead{iflux} & \colhead{ifluxunc} & \colhead{ibkg} & \colhead{flux} & \colhead{fluxunc} & \colhead{bkg}\\ \colhead{ } & \colhead{ } & \colhead{ } & \colhead{ } & \colhead{ } & \colhead{$\mathrm{MJD}$} & \colhead{$\mathrm{DN\,s^{-1}}$} & \colhead{$\mathrm{DN}$} & \colhead{$\mathrm{s}$} & \colhead{$\mathrm{DN\,s^{-1}}$} & \colhead{$\mathrm{DN\,s^{-1}}$} & \colhead{$\mathrm{DN\,s^{-1}}$} & \colhead{$\mathrm{mJy}$} & \colhead{$\mathrm{mJy}$} & \colhead{$\mathrm{MJy\,sr^{-1}}$}}
\startdata
16 Cyg B & 1538 & SolarAnalogs & F1130W & SUB64 & 59762.5 & 5.17e+04 & 4.40e+04 & 0.85 & 8.79e+05 & 1.07e+03 & 1.43e+02 & 4.14e+02 & 5.03e-01 & 1.61e+02 \\
16 Cyg B & 1538 & SolarAnalogs & F1280W & SUB64 & 59762.5 & 9.02e+04 & 4.61e+04 & 0.51 & 1.80e+06 & 2.45e+03 & 4.08e+02 & 3.26e+02 & 4.43e-01 & 1.74e+02 \\
16 Cyg B & 1538 & SolarAnalogs & F1500W & SUB64 & 59762.5 & 5.91e+04 & 4.02e+04 & 0.68 & 1.49e+06 & 4.31e+03 & 6.57e+02 & 2.39e+02 & 6.91e-01 & 2.46e+02 \\
16 Cyg B & 1538 & SolarAnalogs & F1800W & SUB64 & 59762.5 & 2.55e+04 & 2.17e+04 & 0.85 & 7.98e+05 & 2.34e+03 & 5.01e+02 & 1.69e+02 & 4.95e-01 & 2.46e+02 \\
16 Cyg B & 1538 & SolarAnalogs & F2100W & SUB64 & 59762.5 & 1.80e+04 & 1.54e+04 & 0.85 & 7.33e+05 & 1.11e+03 & 6.32e+02 & 1.28e+02 & 1.93e-01 & 2.59e+02 \\
16 Cyg B & 1538 & SolarAnalogs & F2550W & SUB64 & 59762.5 & 5.23e+03 & 1.34e+04 & 2.55 & 2.47e+05 & 6.08e+02 & 1.11e+03 & 8.12e+01 & 2.00e-01 & 8.51e+02 \\
2MASS J17430448+6655015 & 1027 & ADwarfs & F560W & FULL & 59737.6 & 6.68e+02 & 1.85e+04 & 27.75 & 4.76e+03 & 7.35e+00 & 3.19e+00 & 8.78e-01 & 1.36e-03 & 1.43e+00 \\
2MASS J17430448+6655015 & 1027 & ADwarfs & F560W & FULL & 59724.4 & 3.91e+02 & 1.09e+04 & 27.75 & 3.38e+03 & 7.09e+00 & 3.43e+00 & 6.24e-01 & 1.31e-03 & 1.54e+00 \\
2MASS J17430448+6655015 & 1536 & ADwarfs & F560W & FULL & 59763.0 & 5.92e+02 & 1.64e+04 & 27.75 & 4.71e+03 & 7.10e+00 & 3.28e+00 & 8.69e-01 & 1.31e-03 & 1.47e+00 \\
2MASS J17430448+6655015 & 1027 & ADwarfs & F560W & FULL & 59727.9 & 6.52e+02 & 1.81e+04 & 27.75 & 4.78e+03 & 5.33e+00 & 3.23e+00 & 8.83e-01 & 9.84e-04 & 1.45e+00 \\
2MASS J17430448+6655015 & 1027 & ADwarfs & F770W & FULL & 59727.9 & 5.64e+02 & 1.57e+04 & 27.75 & 4.69e+03 & 6.72e+00 & 1.86e+01 & 4.99e-01 & 7.15e-04 & 4.79e+00 \\
2MASS J17430448+6655015 & 1536 & ADwarfs & F770W & FULL & 59763.0 & 5.18e+02 & 1.44e+04 & 27.75 & 4.67e+03 & 6.37e+00 & 1.70e+01 & 4.97e-01 & 6.78e-04 & 4.38e+00 \\
2MASS J17430448+6655015 & 1027 & ADwarfs & F770W & FULL & 59737.6 & 5.18e+02 & 1.44e+04 & 27.75 & 4.66e+03 & 6.38e+00 & 1.75e+01 & 4.95e-01 & 6.78e-04 & 4.51e+00 \\
2MASS J17430448+6655015 & 1536 & ADwarfs & F1000W & FULL & 59763.1 & 1.99e+02 & 5.53e+03 & 27.75 & 2.03e+03 & 4.58e+00 & 3.46e+01 & 2.92e-01 & 6.58e-04 & 1.20e+01 \\
2MASS J17430448+6655015 & 1027 & ADwarfs & F1000W & FULL & 59737.6 & 2.01e+02 & 5.56e+03 & 27.75 & 2.05e+03 & 4.68e+00 & 3.56e+01 & 2.94e-01 & 6.72e-04 & 1.23e+01 \\
2MASS J17430448+6655015 & 1027 & ADwarfs & F1000W & FULL & 59727.9 & 2.10e+02 & 5.84e+03 & 27.75 & 2.04e+03 & 4.77e+00 & 3.75e+01 & 2.92e-01 & 6.85e-04 & 1.30e+01 \\
2MASS J17430448+6655015 & 1536 & ADwarfs & F1130W & FULL & 59763.1 & 4.84e+01 & 2.68e+03 & 55.50 & 5.03e+02 & 1.84e+00 & 1.56e+01 & 2.29e-01 & 8.39e-04 & 1.69e+01 \\
2MASS J17430448+6655015 & 1027 & ADwarfs & F1130W & FULL & 59737.6 & 5.07e+01 & 2.81e+03 & 55.50 & 5.02e+02 & 2.06e+00 & 1.60e+01 & 2.29e-01 & 9.39e-04 & 1.73e+01 \\
2MASS J17430448+6655015 & 1027 & ADwarfs & F1130W & FULL & 59727.9 & 5.13e+01 & 2.85e+03 & 55.50 & 5.00e+02 & 1.89e+00 & 1.67e+01 & 2.27e-01 & 8.61e-04 & 1.81e+01 \\
2MASS J17430448+6655015 & 1536 & ADwarfs & F1280W & FULL & 59763.1 & 1.11e+02 & 4.91e+03 & 44.40 & 1.01e+03 & 4.34e+00 & 5.36e+01 & 1.76e-01 & 7.59e-04 & 2.21e+01 \\
2MASS J17430448+6655015 & 1027 & ADwarfs & F1280W & FULL & 59727.9 & 1.14e+02 & 5.08e+03 & 44.40 & 1.02e+03 & 4.57e+00 & 5.68e+01 & 1.78e-01 & 7.95e-04 & 2.33e+01 \\
2MASS J17430448+6655015 & 1027 & ADwarfs & F1280W & FULL & 59737.6 & 1.11e+02 & 4.94e+03 & 44.40 & 1.02e+03 & 4.37e+00 & 5.49e+01 & 1.77e-01 & 7.61e-04 & 2.25e+01 \\
2MASS J17430448+6655015 & 1027 & ADwarfs & F1500W & FULL & 59737.6 & 1.44e+02 & 3.44e+04 & 238.90 & 8.47e+02 & 3.02e+00 & 1.09e+02 & 1.31e-01 & 4.67e-04 & 3.91e+01 \\
2MASS J17430448+6655015 & 1027 & ADwarfs & F1500W & FULL & 59727.9 & 1.47e+02 & 3.51e+04 & 238.90 & 8.66e+02 & 2.60e+00 & 1.11e+02 & 1.34e-01 & 4.01e-04 & 3.99e+01 \\
2MASS J17430448+6655015 & 1536 & ADwarfs & F1500W & FULL & 59763.1 & 1.44e+02 & 7.21e+03 & 49.95 & 8.44e+02 & 4.17e+00 & 1.08e+02 & 1.31e-01 & 6.46e-04 & 3.89e+01 \\
G 191-B2B & 1537 & HotStars & F560W & FULL & 59830.4 & 6.85e+02 & 1.90e+04 & 27.75 & 4.39e+03 & 6.49e+00 & 4.81e+00 & 8.11e-01 & 1.20e-03 & 2.16e+00 \\
G 191-B2B & 1537 & HotStars & F770W & FULL & 59830.4 & 5.33e+02 & 1.48e+04 & 27.75 & 4.30e+03 & 6.47e+00 & 3.10e+01 & 4.58e-01 & 6.88e-04 & 7.99e+00 \\
G 191-B2B & 1537 & HotStars & F1000W & FULL & 59830.4 & 2.07e+02 & 5.74e+03 & 27.75 & 1.91e+03 & 5.37e+00 & 5.72e+01 & 2.75e-01 & 7.75e-04 & 1.98e+01 \\
G 191-B2B & 1537 & HotStars & F1130W & FULL & 59830.4 & 5.73e+01 & 3.82e+03 & 66.60 & 4.72e+02 & 2.12e+00 & 2.59e+01 & 2.15e-01 & 9.65e-04 & 2.82e+01 \\
G 191-B2B & 1537 & HotStars & F1280W & FULL & 59830.4 & 1.36e+02 & 6.06e+03 & 44.40 & 9.71e+02 & 5.08e+00 & 8.14e+01 & 1.71e-01 & 8.95e-04 & 3.38e+01 \\
G 191-B2B & 1537 & HotStars & F1500W & FULL & 59830.4 & 1.77e+02 & 1.18e+04 & 66.60 & 8.29e+02 & 4.30e+00 & 1.42e+02 & 1.29e-01 & 6.70e-04 & 5.17e+01 \\
\enddata
\tablecomments{Table~\ref{tab:phot} is published in its entirety in the machine-readable format. A portion is shown here for guidance regarding its form and content.}
\end{deluxetable*}
\end{rotatetable*}

\subsection{Photometry}
\label{sec:phot}

\begin{figure}[tbp]
\epsscale{1.2}
\plotone{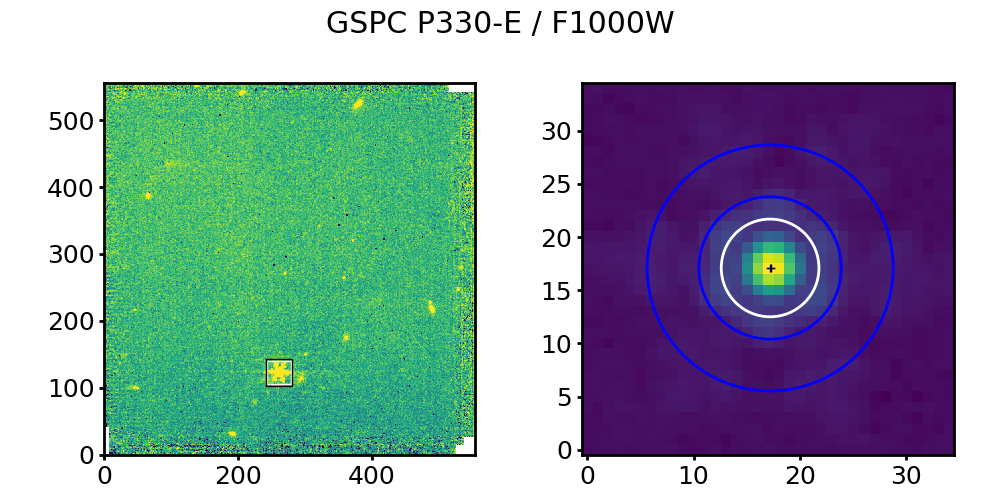}
\caption{A visualization of the photometric method for GSPC P330E observed with the F1000W imaging band with the BRIGHTSKY subarray.
The full mosaic is shown on the left with a white and black box surrounding the target.
The right panel zooms in on the target region with the aperture in white, the sky annulus in blue, and the final centroid position marked with a cross.
The x and y axis units are pixels, which are $0.11\arcsec$ across.
The minima and maxima of the color scales differ between the two images.
\label{fig:phot_example}}
\end{figure}

The photometry for each observation was measured using the photutils package \citep{Bradley22} from the mosaics using the apertures and aperture corrections from Table~\ref{tab:apcor}.
The center of the aperture was determined by finding the initial position of the star based on the known coordinates and proper motion, followed by center-of-mass centroiding on the brightest source near this initial position.
The photometry accounts for partial pixels in the circular aperture.
The uncertainty in the photometry was calculated based on the pipeline propagated uncertainties and includes a second term that gives the uncertainty in the background subtraction based on the measured scatter in the background. 
A visual example of the photometry is shown in Fig.~\ref{fig:phot_example}.
The photometry in this fixed aperture was then corrected to infinite aperture using the appropriate aperture correction.  
Table~\ref{tab:phot} gives the resulting photometry along with additional observational details.
This table includes the star name (name), program identification (PID), the source type (srctype), the filter, the subarray, the observation time (time), the central-pixel rate (pixrate), the central-pixel well depth (pixwelldepth), the integration time (inttime), the instrumental flux (iflux), instrumental flux uncertainty (ifluxunc), the instrumental average background per pixel (ibkg), and the equivalent of the last three in physical units (flux, fluxunc, and bkg) using the calibration from section~\ref{sec:calfacs}.

\subsection{Model Predictions}
\label{sec:models}

The predicted flux densities for the hot stars and A dwarfs were determined using CALSPEC\footnote{\url{https://www.stsci.edu/hst/instrumentation/reference-data-for-calibration-and-tools/astronomical-catalogs/calspec}} models \citep{Bohlin14, Bohlin22}.
For the Solar analogs, models by \citet{Rieke24} were used.
These models are the combination of stellar atmosphere models extinguished by dust.
See \citet{Gordon22} for the stellar and dust extinction properties of each star.
The band flux densities were calculated using the MIRI band-response functions provided by Pandeia \citep{Pontoppidan16}.
The band flux densities were calculated in $F(\lambda)$ units using Equation~5 from \citet{Gordon22} and converted to $F(\nu)$ flux densities using the pivot wavelength $\lambda_\mathrm{ref}$ with the standard conversion \citep[e.g., Equation~11 of][]{Bohlin14}.
This step supports the two common photometric conventions \citep{Gordon22}.

\subsection{Zero-Magnitude Flux Densities in the Sirius-Vega System}

\begin{deluxetable}{lcc}[tbp]
\tablewidth{0pt}
\tabletypesize{\small}
\tablecaption{Zero-Magnitude Flux Densities in the Sirius-Vega System\label{tab:siriusvega}}
\tablehead{\colhead{Band} & \colhead{$F(\lambda)$} & \colhead{$F(\nu)$} \\
 & \colhead{(ergs \AA$^{-1}$ cm$^{-2}$ s$^{-1}$)} & \colhead{(Jy)}}
\startdata
F560W & 1.090e-12 & 115.439 \\
F770W & 3.340e-13 & 65.011 \\
F1000W & 1.162e-13 & 38.401 \\
F1065C & 8.963e-14 & 33.562 \\
F1140C & 6.929e-14 & 29.534 \\
F1130W & 6.924e-14 & 29.538 \\
F1280W & 4.269e-14 & 23.368 \\
FND & 5.175e-14 & 28.724 \\
F1500W & 2.249e-14 & 17.020 \\
F1550C & 1.976e-14 & 15.858 \\
F1800W & 1.103e-14 & 11.895 \\
F2100W & 6.227e-15 & 8.981 \\
F2300C & 4.383e-15 & 7.517 \\
F2550W & 2.802e-15 & 6.012 \\
\enddata
\end{deluxetable}

Table~\ref{tab:siriusvega} provides the flux densities for zero magnitude in the Sirius-Vega system computed in $F(\lambda)$ units as defined for the JWST band fluxes \citep{Gordon22} and the $F(\nu)$ values computed as discussed by \citet{Bohlin14}.
Magnitudes are not required for the absolute flux calibration, but are used extensively in many studies.
They are provided here as a convenience and are based on the known filter bandpasses and the model for Sirius shifted to the brightness of Vega.
This approach uses Sirius as the color standard and preserves the historical zero magnitude definition based on Vega \citep{Rieke22}.
The definition of the JWST band fluxes supports both photometric conventions \citep[see][]{Gordon22}.
For the photometric convention that uses color corrections see the appropriate JDox page\footnote{To be added to JDox very soon} for corrections for a selection of spectral shapes.

\section{Results \label{sec:results}}

Following \citet{Gordon22}, the flux calibration factor that converts instrumental DN~s$^{-1}$~pixel$^{-1}$ to physical surface brightnesses in MJy~sr$^{-1}$ is
\begin{equation}
C = \frac{F}{N_\mathrm{ap} A_\mathrm{cor} \Omega_{\mathrm{pix}}} .
\end{equation}
where $F$ is the model flux density in MJy (\S\ref{sec:models}), $N$ is the measured flux density in DN~s$^{-1}$~pixel$^{-1}$ in a finite aperture (iflux in Table~\ref{tab:phot}), $A_\mathrm{cor}$ is the aperture correction to an infinite aperture (\S\ref{sec:apcor}), and $\Omega_{pix}$ is the average solid angle per pixel (in sr).

Ideally, the flux calibration factor for each band would be a single constant converting instrumental to physical units.
As part of this analysis, we have found that the flux calibration depends on both the time of observation and the subarray used.
Section~\ref{sec:repeat} and \ref{sec:subtrans} discuss the dependence on time and subarray, respectively.
We did not find that the flux calibration depends significantly on other properties of the detectors or calibration stars.
The source parameters investigated include the predicted flux densities, the type of star (hot stars, A dwarfs, and Solar analogs), and measured backgrounds (\S\ref{sec:srcdep}).
The detector parameters investigated include the central-pixel well depth in DN, the instrumental central-pixel rate in DN~s$^{-1}$, and integration time (\S\ref{sec:detector}). 
Section~\ref{sec:calfacs} quantifies the time- and subarray-dependent flux calibration and accuracy.

\subsection{Temporal Dependence}
\label{sec:repeat}

\begin{figure*}[tbp]
\epsscale{1.2}
\plotone{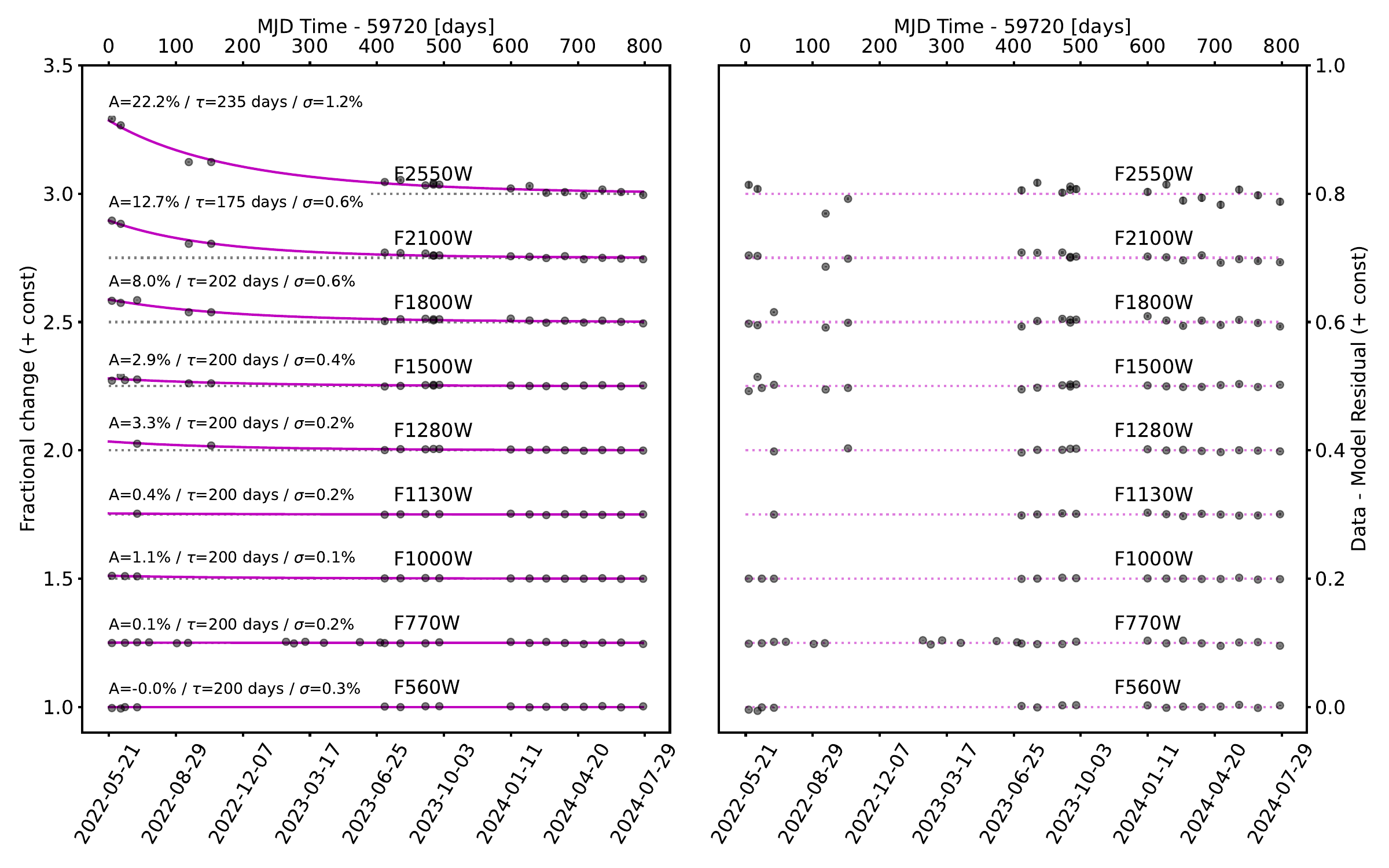}
\caption{Left: The time-dependent behavior of the repeated observations of BD+60~1753 plus two other stars for all the imaging filters. 
An exponential model fit to the observations in each band is plotted in this panel as the solid purple line.
The dotted line gives the value after the modeled exponential has fully decayed. 
The amplitude $A$, the time constant $\tau$, and standard deviation $\sigma$ around the exponential fit are given on the left.
Right: The residuals after subtraction of the exponential model.
\label{fig:repeat}}
\end{figure*}

The repeatability of a source was monitored through monthly observations of BD+60~1753, an A dwarf that is relatively bright and is known to have very low short-term intrinsic variability \citep[$<$0.03\%, ][]{Gordon22}.
The objective is to measure the intrinsic scatter of observations due to variability of the telescope (e.g., PSF) and instrument (e.g., detectors).
Focusing on the F770W observations (Fig.~\ref{fig:repeat}), the repeatability is measured to be 0.2\%, comparable to that measured for the Spitzer/MIPS 24~\micron\ band \citep{Engelbracht07} and lower than the 1--2\% measured for the IRAC bands \citep{Reach05, Bohlin11, Krick21} .

After the detection of the significant response change at long wavelengths for the MIRI Medium Resolution Spectrometer (MRS) \citep{Law24}, these observations were expanded to include all imaging filters every month\footnote{\url{https://www.stsci.edu/contents/news/jwst/2023/temporal-behavior-of-the-miri-reduced-count-rate}}.
With this change, the goal of the monitoring of BD+60~1753 expanded to include detecting and measuring any temporal changes in total system response.
Combining these expanded observations with observations of the same star taken during JWST Commissioning led to the discovery that the long-wavelength imaging filters also suffer from a temporal degradation in transmission.
Fig.~\ref{fig:repeat} shows the relative change in the flux of BD+60~1753 for all 9 imaging bands.
In addition, pairs of measurements of $\delta$~UMi and HD~37962 are included to add critical measurements in the gap between the end of Commissioning and the start in cycle 2 of the expanded, all-filter observations of BD+60~1753.
Both of these additional stars had been observed in multiple long-wavelength filters between days 100 and 200.
Repeated observations between days 475 and 500 were taken of both stars to provide a relative measurement.
This allowed the earlier measurements to be included in the repeatability measurement by normalizing the latter observations to observations of BD+60~1753 close in time.

The F2550W filter shows a clear degradation in total system responsivity of $\sim$22\%.
The degradation can be seen in other filters, with the amplitude decreasing monotonically with wavelength to 3\% in F1280W.
For the shorter-wavelength filters, the degradation is small and generally consistent with very small or no degradation.
The imaging response degradation at F2550W of $\sim$22\% is significantly smaller than the MRS degradation seen at the equivalent wavelengths \citep{Law24}.
The impact on the S/N of an observation with F2550W is $\sim$11\%.

The expanded monitoring of BD+60~1753 measures the repeatability of observations in all imaging filters.
As Fig.~\ref{fig:repeat} shows, discrepancies in the repeatability amplitude are lowest for the short-wavelength bands with values of 0.1--0.4\% and grow to 1.2\% for the F2550W band.
These uncertainties set the maximum S/N possible from a single, dithered observation.

\subsection{Anomalous Stars/Observations}
\label{sec:stars_removed}

One of the motivations for observing a sample of stars is to identify stars that are not suitable for absolute flux calibration.
We found one star that fell into this category.
In addition, a few observations in specific filters were found to be significantly deviant from the average.

The A dwarf HD~180609 shows a clear signature of a debris disk, with an observed excess above the model prediction for long-wavelength bands that increases with wavelength.
The signature of such an excess is a decreasing calibration factor as a function of wavelength, and this is clearly seen for this star.
This makes this star unsuitable for flux calibration measurements and it was removed from all analyses except for the subarray dependence as it is the only star observed in all filters with SUB128 (see below).
Debris disks have appeared in samples of presumably well-vetted standard stars before.
In the Spitzer \citep{Werner04} Infrared Spectrometer \citep{Houck04} absolute flux program, four of the 20 A dwarf standards had a clear debris disk, plus three suspects, along with one of the 30 K giant standards \citep{Sloan04, Sloan15}.
In the Spitzer Mid-Infrared Instrument \citep{Rieke04} absolute flux program, one A dwarf and one K giant were found to have the signature of a debris disk \citep{Gordon07}.
The difficulty of avoiding these occasional anomalies drives the need for observing a sample of stars of each stellar type.

As part of computing the calibration factors, some observations were found to be significantly different than average.
Often these deviant observations were seen for a single observation for a star, with all the other observations for the same star being consistent with the average.
The deviant observations are indicated in the calibration factor plots in the rest of this paper.
There was one star where all its observations were flagged as deviant.
This was C26202 that was observed in F560W, F770W, and F1000W and the deviations seen were both above and below the average.
The reason for all the observations being deviant is under investigation.

\subsection{Subarray Dependence}
\label{sec:subtrans}

\begin{figure}[tbp]
\epsscale{1.2}
\plotone{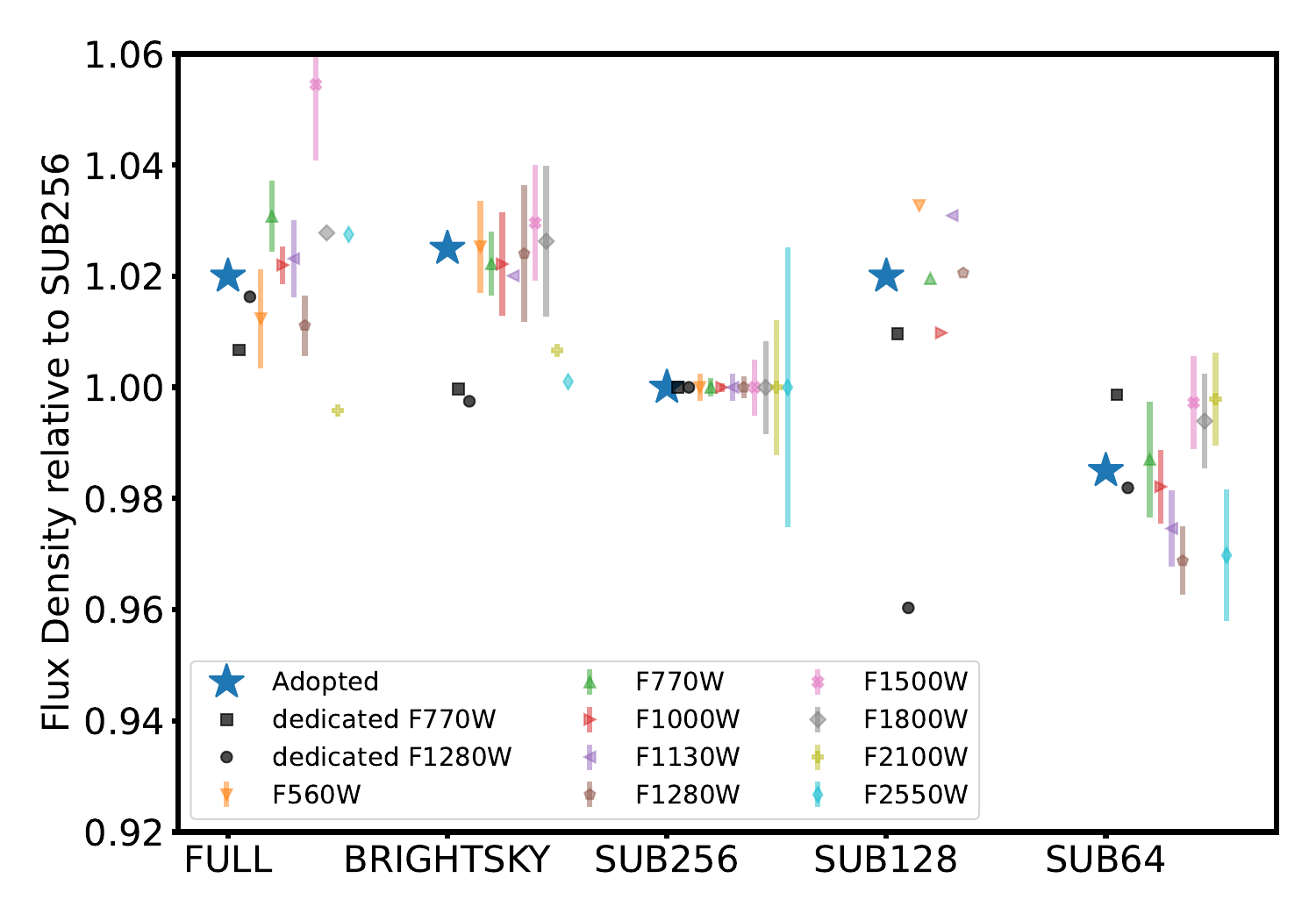}
\caption{The fractional change in flux density relative to SUB256 for each subarray using a variety of observations.
The dedicated subarray transfer observations taken in F770W and F1280W are simple ratios of the measured flux densities where multiple measurements in the same subarray were averaged if available.
All of the flux-calibration observations were used to measure equivalent ratios using the weighted average calibration factors for each subarray and band.
The uncertainties for these ratios are the weighted standard deviation of the mean.
No error bars are shown for three or fewer observations.
The blue stars are the adopted relative subarray throughputs.
\label{fig:subtrans}}
\end{figure}

\begin{deluxetable}{lc}[tbp]
\tablewidth{0pt}
\tabletypesize{\small}
\tablecaption{Subarray Dependence\label{tab:subtrans}}
\tablehead{\colhead{Subarray} & \colhead{$D_\mathrm{SA}$}}
\startdata
FULL & 1.000 \\
BRIGHTSKY & 1.005 \\
SUB256 & 0.980 \\
SUB128 & 1.000 \\
SUB64 & 0.966 \\
\enddata
\end{deluxetable}

Observations for absolute flux calibration are taken with different subarrays so that both faint and bright targets can be observed.
The MIRI Imager has five subarrays including FULL using all the detector pixels and four smaller subarrays (BRIGHTSKY, SUB256, SUB128, SUB64) with each successively reducing the area read by a factor of four \citep{Ressler15}.
Measuring potential differences in the observed flux densities in different subarrays is important to ensuring a consistent calibration.
Possible causes of variations between subarrays include different readout patterns, transients associated with switching subarrays, and flatfield spatial errors as the subarrays are in different detector locations \citep{Gillian23, Dicken24}.

Dedicated observations to measure variations in the flux calibration with subarray were taken in two filters with two different stars.
2MASS J17571324+6703409 was observed back-to-back in the F770W band for all five subarrays.
2MASS~J18022716+6043356 was observed back-to-back in the F1280W band for all five subarrays, with the FULL and SUB64 subarrays observed multiple times.
For the F1280W measurements, the FULL frame observations were taken at the beginning and end of the observation sequence to check that any changes seen did not depend on the order of the observations.
Fig.~\ref{fig:subtrans} shows the flux density in each subarray relative to that observed in the SUB256 subarray.
For all but the SUB128 subarray, these two sets of measurements agree well.
For the SUB128 subarray, they differ by $\sim$4\%, but the cause is not yet known, and additional dedicated observations will be needed to determine it.

The rest of the observations can be used to probe the subarray dependence as well.
First, the weighted average calibration factor and uncertainty for each subarray in each band is calculated after correcting for the known temporal dependence.
The weighting was such that each star had equal weight in the average (e.g., a star with $n$ observations has its observations each given a weight of $n^{-1}$).
The equivalent measurement can then be computed by taking the ratio of the SUB256 subarray calibration factor to that measured in the other subarrays.
This ratio has the SUB256 calibration factor in the numerator as the measured flux densities are in the denominator of the calibration factor calculation.
The SUB256 subarray was used as the reference as it was the subarray with the most observations with flux calibration data in all bands.

Fig.~\ref{fig:subtrans} shows the resulting ratios.
For FULL and SUB64, these ratios agree well with those measured from the dedicated observations.
For BRIGHTSKY, the ratios disagree for both the F770W and F1280W dedicated measurements when considering the averages based on $>$3 observations.
For SUB128, the ratios agree well with the dedicated F770W observations, but not those taken with F1280W.
The SUB128 ratio values are dominated by HD~180609, a star with a clear signature of a debris disk in F1500W and longer bands (see Section~\ref{sec:stars_removed}).
We have included the F1280W and shorter bands for this star in the analysis only for the subarray dependence because it is the only star observed in SUB128 for all bands.
In the end, the SUB128 dependence is the same as the FULL subarray, resulting in no SUB128 correction compared to FULL.

In summary, the dedicated subarray observations for the SUB128 subarray do not agree and these dedicated observations do not agree with the ratios based on calibration factors for the BRIGHTSKY subarray.
The two sets of observations are consistent enough that we can conclude that the subarrays differ significantly.
Visually examining Fig.~\ref{fig:subtrans}, we have adopted subarray dependencies mainly influenced by the calibration factor ratios, but also taking into account the dedicated observations.
The calibration factor ratios are preferred because (1) they are based on many more observations than the dedicated subarray observations, and (2) the dedicated subarray observations could be impacted by subarray switching transients.
The subarray was measured relative to SUB256, because SUB256 was well measured in all bands.
It is preferable to have the subarray dependence relative to the FULL subarray that is used for most MIRI imaging observations.
Hence, the $D_\mathrm{SA}$ dependence as given in Table~\ref{tab:subtrans} is given relative to FULL frame by dividing the adopted values shown in Fig.~\ref{fig:subtrans} by the adopted FULL value.
Thus, $D_\mathrm{SA}$ gives the throughput for each subarray relative to the FULL frame.

\subsection{Source Dependencies}
\label{sec:srcdep}

\begin{figure*}[tbp]
\epsscale{1.2}
\plotone{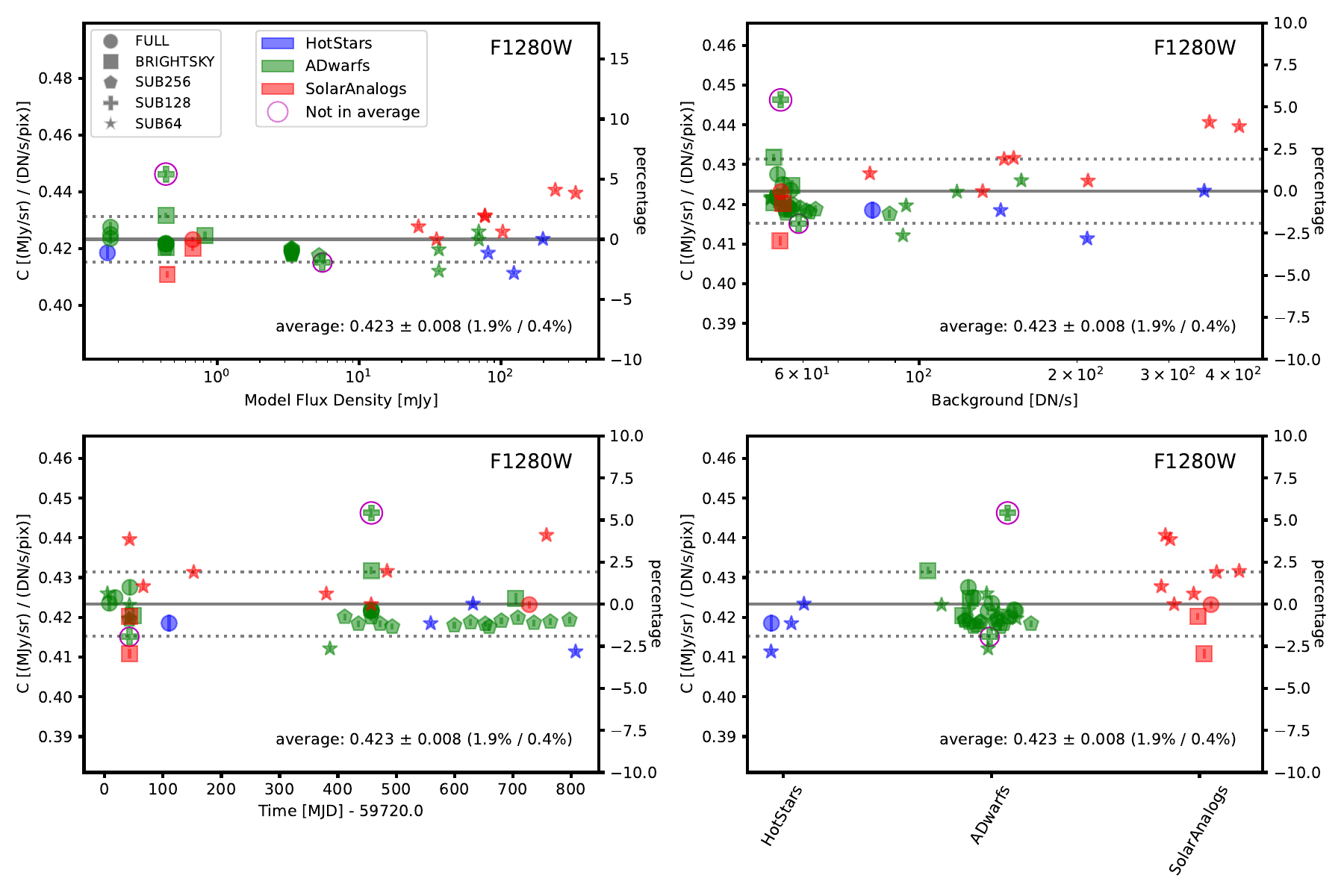}
\caption{The F1280W calibration factors plotted as a function of the source predicted flux density, measured background, observation time, and source type.
The calibration factors are corrected for the time and subarray dependencies.
Each point represents one observation for one star.
The colors and symbols to distinguish different types of stars and subarrays are the same in all panels.
While the observation of HD~180609 is within the scatter for this band, it is flagged and not used (\S\ref{sec:stars_removed}).
In addition, one observation is flagged as $> 3.5\sigma$ from the weighted average and not used (\S\ref{sec:stars_removed}). 
The solid horizontal gray line gives the weighted average in each panel, with the $\pm 1\sigma$ standard deviations as dotted gray lines.
The average calibration factors and uncertainties are given in the lower right of each plot.
The percentage standard deviation and standard deviation of the mean are given in parentheses as well.
\label{fig:f1280w_src}}
\end{figure*}

Fig.~\ref{fig:f1280w_src} shows how the F1280W calibration factors for each observation depend on multiple source parameters.
Investigating if the calibration factor depends on properties of the source is important to test that the calibration factor is applicable to as wide a range of observations as can be tested.
The weighted averages and uncertainties were computed as described in \S\ref{sec:subtrans}.

The plot versus model flux density shows that the calibration is constant for sources from 0.2 to 200~mJy.
The lack of a dependence on flux density in all but one of the bands (see appendix Fig.~\ref{fig:calfacs_mflux}) leads to two conclusions.
First, the MIRI detectors do not suffer from significant non-linearities due to a dependence on flux density \citep{Bohlin06, deJong06}.  
Second, the known brighter-fatter effect \citep{Argyriou23} is not affecting the photometry in the aperture used.
The one possible exception to this result is the FND band, which shows a small trend with predicted flux density (as discussed further in the next subsection).

Another possible dependence could be on the measured background.
The background affects all pixels similarly unlike a point source with its strong gradients in the illumination between pixels.
The plot versus background for F1280W and for all the bands (see appendix Fig.~\ref{fig:calfacs_bkg}) shows no dependence on such uniform illumination.

The temporal dependence determined from repeated observations of mainly one star discussed in \S\ref{sec:repeat} is tested using all the observations in the lower left plot of this figure.
The plot shows that the temporal correction is working well for all observations for F1280W and, in fact, all bands (see appendix Fig.~\ref{fig:calfacs_time}).

\begin{figure}[tbp]
\epsscale{1.2}
\plotone{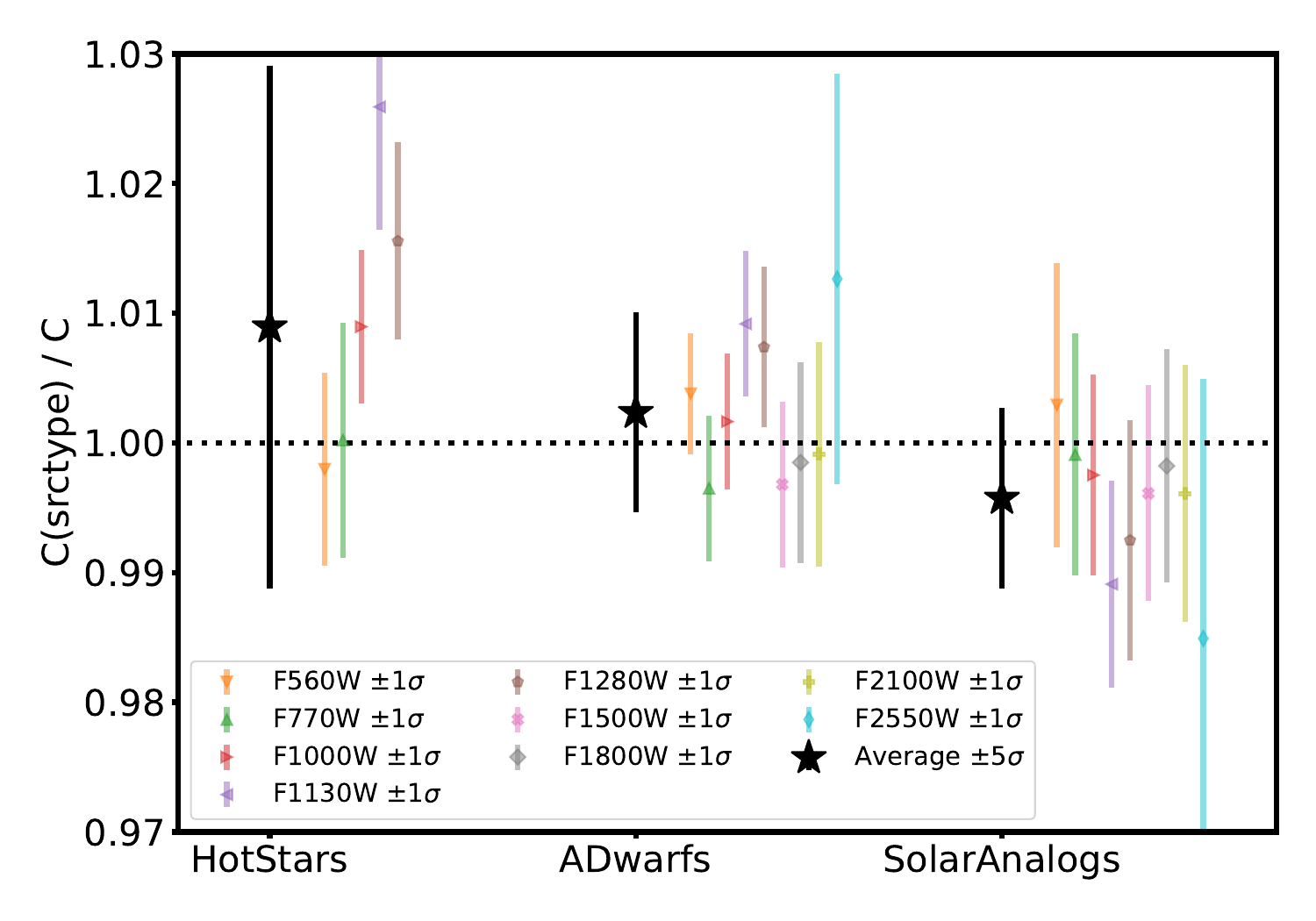}
\caption{The ratio of the average calibration factor for each band for each source type versus the average calibration factor per band for all source types.
Only bands with $> 3$ individual measurements are shown as this is required to measure an uncertainty.
The error bars for the individual bands are one $\sigma$ standard deviations.
Black stars and vertical lines depict the weighted average of all bands and five times the standard deviation of the mean ($5\sigma$).
\label{fig:srctype}}
\end{figure}

The lower right panel in Fig.~\ref{fig:f1280w_src} probes the dependence on the type of calibration star.
No obvious dependence on type of star is found for F1280W or for the other bands (see appendix Fig.~\ref{fig:calfacs_srctype}).
To quantify the difference between source types, the average calibration factors for each band were computed by source type.
Fig.~\ref{fig:srctype} plots the ratios between these values and the average calibration factors per band for all source types (in colors), along with the weighted averages of all bands per source type (in black) and error bars that represent the $5\sigma$ standard deviation.
The averages for all the types of stars are within $5\sigma$ of unity, indicating no significant difference in the calibration factors between types.
These results indicate that the calibration does not depend systematically on source type.

\subsection{Detector Dependencies}
\label{sec:detector}

\begin{figure*}[tbp]
\epsscale{1.2}
\plotone{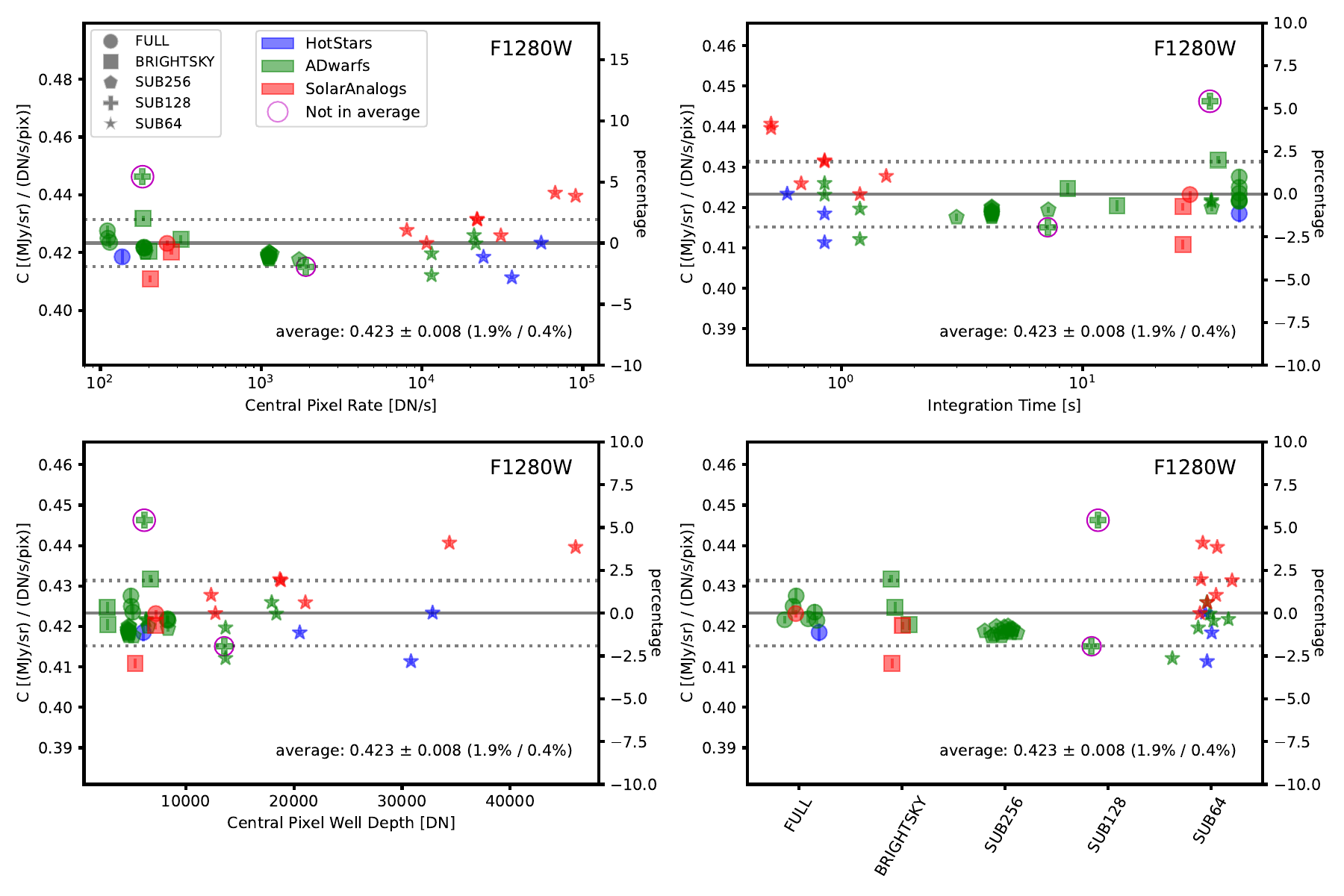}
\caption{The calibration factor versus the detector parameters central-pixel rate, integration time, central-pixel well depth, and subarray.
The calibration factors plotted have been corrected for the temporal and subarray dependencies.
Different colors and symbols indicate different types of stars and subarrays.
While the observation of HD~180609 is within the scatter for this band, it is flagged and not used (\S\ref{sec:stars_removed}).
In addition, one observation is flagged as $> 3.5\sigma$ from the weighted average and not used (\S\ref{sec:stars_removed}). 
The horizontal gray solid line shows weighted average, with dotted lines for the $\pm 1\sigma$ standard deviation.
The calibration factors and uncertainties are given in the lower right of each plot.
The percentage standard deviation and standard deviation of the mean are in parentheses as well.
\label{fig:f1280w_det}}
\end{figure*}

The calibration factor can depend on detector parameters as has already been shown for subarrays.
Fig.~\ref{fig:f1280w_det} plots the F1280W calibration factors after correction for the time and subarray dependence as a function of the detector parameters central-pixel rate, central-pixel well depth, integration time, and subarray.
The weighted averages and uncertainties were computed as described in \S\ref{sec:subtrans}.

The dependence on the rate of arriving photons is investigated in the upper left of Fig.~\ref{fig:f1280w_det}.
This panel closely resembles the plot in the previous section versus model flux, except that the central-pixel rate is used given in instrumental DN~s$^{-1}$.
As in Fig.~\ref{fig:f1280w_src}, no clear dependence is seen for F1280W or any of the other filters except for the FND filter (see appendix Fig.~\ref{fig:calfacs_rate}).

The central-pixel well depth probes the accuracy of the applied non-linearity correction.
The lower left panel in Fig.~\ref{fig:f1280w_det} shows no clear dependence on central-pixel well depth, indicating that the non-linearity correction is quite accurate for F1280W and all bands except the FND (see appendix Fig.~\ref{fig:calfacs_welldepth}).

The FND band shows a trend with predicted flux density, central-pixel rate, and central-pixel well depth.
This result is surprising, because all of the wavelengths covered by the FND band are covered with other filters that do not show such trends.
A possible origin of the FND behavior may be the result of its very wide bandpass, with a width of 6.7~\micron\ centered on 12.9~\micron. 
The MIRI non-linearity correction is known to depend on wavelength \citep{Morrison23}, with longer-wavelength bands having a different non-linearity that shorter-wavelength bands.
The FND observations were processed using the non-linearity correction for bands F1280W and shorter, but they may possibly have a different non-linearity as the red end of the band extends well beyond the F1280W band.
Future non-linearity work will investigate this possible explanation.

The upper right panel in Fig.~\ref{fig:f1280w_det} probes the dependence of calibration factor on the integration time, which could be caused by reset-induced transients that occur near the beginning of a ramp and could impact shorter ramps more than longer ramps.
No clear dependence on integration times from 0.5 to 40~s for F1280W is found.
When including all the bands (see appendix Fig.~\ref{fig:calfacs_inttime}), no significant dependence is seen for integrations times from 0.5 to near 300~s.
This lack of dependencies indicates that the reset transients are not significantly affecting the calibration factors.

Finally, the accuracy of the subarray dependence derived in \S\ref{sec:subtrans} is tested for F1280W in the lower right panel of Fig.~\ref{fig:f1280w_det}.
This plot clearly shows that the subarray correction is accurate, which is not surprising as it was derived partially with these observations.
This result is also true for all of the other bands, as can be seen in the appendix Fig.~\ref{fig:calfacs_subarr}.
The lack of a sufficient number of observations prevented a similar check for the SUB128 subarray.
The only star observed in all bands with SUB128 has a debris disk (\S\ref{sec:stars_removed}).

\subsection{Calibration Factors}
\label{sec:calfacs}

\begin{deluxetable*}{lcccccccc}[tbp]
\tablewidth{0pt}
\tablecaption{Calibration Factor Info\label{tab:calfac}}
\tablehead{\colhead{Band} & \colhead{$A$\tablenotemark{$\star$}} & \colhead{$B$\tablenotemark{$\star$}} & \colhead{$B$} & \colhead{$\tau$} & \colhead{$\sigma(CF)$\tablenotemark{$\star$}} & \colhead{$\sigma(CF)$} & \colhead{$n_\mathrm{stars}$} & \colhead{$\sigma(\mathrm{repeat})$} \\
& & & \colhead{(\%)} & \colhead{(days)} & & \colhead{(\%)} & & \colhead{(\%)}} 
\startdata
\multicolumn{9}{c}{Imaging} \\ \hline
F560W & 0.4496 & 0.0000 & 0.0 & 200.0 & 0.00166 & 0.37 & 11.75 & 0.27 \\ 
F770W & 0.2579 & -0.0001 & 0.1 & 200.0 & 0.00096 & 0.37 & 15.00 & 0.25 \\ 
F1000W & 0.3488 & -0.0038 & 1.1 & 200.0 & 0.00111 & 0.32 & 20.00 & 0.08 \\ 
F1130W & 1.0898 & -0.0042 & 0.4 & 200.0 & 0.00520 & 0.48 & 19.00 & 0.16 \\ 
F1280W & 0.4233 & -0.0140 & 3.3 & 200.0 & 0.00185 & 0.44 & 18.89 & 0.20 \\ 
FND & 41.5591 & -1.3650 & 3.3 & 200.0 & 0.39411 & 0.95 & 10.00 & \nodata \\ 
F1500W & 0.3703 & -0.0107 & 2.9 & 200.0 & 0.00176 & 0.48 & 21.00 & 0.45 \\ 
F1800W & 0.5067 & -0.0407 & 8.0 & 201.7 & 0.00300 & 0.59 & 20.00 & 0.60 \\ 
F2100W & 0.4389 & -0.0557 & 12.7 & 175.0 & 0.00290 & 0.66 & 17.00 & 0.59 \\ 
F2550W & 0.9064 & -0.2011 & 22.2 & 235.4 & 0.00888 & 0.98 & 16.00 & 1.20 \\ 
\hline \multicolumn{9}{c}{Coronagraphy} \\ \hline
F1065C & 3.1125 & -0.0234 & 0.8 & 200.0 & 0.01313 & 0.42 & 5.00 & \nodata \\ 
F1140C & 2.8703 & -0.0112 & 0.4 & 200.0 & 0.02100 & 0.73 & 5.00 & \nodata \\ 
F1550C & 3.8974 & -0.1441 & 3.7 & 200.3 & 0.02841 & 0.73 & 4.00 & \nodata \\ 
F2300C & 1.1106 & -0.1852 & 16.7 & 200.4 & 0.00996 & 0.90 & 4.00 & \nodata \\ 
\enddata
\tablenotetext{\star}{Units of (MJy sr$^{-1}$) / (DN s$^{-1}$ pixel$^{-1}$).}
\end{deluxetable*}

The average calibration factors are computed after correcting for the dependencies on time and subarray so that all measurements could be used.
Averaging measurements of calibration factors made for different stars results in a more accurate flux calibration.
A single star has been found empirically to be modeled to an accuracy of $\sim$2\% \citep{Bohlin14} and, hence, to improve on this accuracy, multiple stars need to be averaged.
In addition, any single star can have properties that make it systematically challenging to model.
The debris disk found as part of this work is one example.
The most famous case of such confounding issues is for Vega that is so rapidly rotating that it has a large gradient in its stellar temperature from pole to equator {\em and} it has a debris disk clearly seen in the near- and mid-IR \citep{Aumann84, Su05, Aufdenberg06}.
Hence, observing multiple stars allows for (1) the removal of such confounding sources and (2) averaging the remainder to improve the flux-calibration accuracy.

Combining the time and subarray dependencies with the average factors, the calibration factor as a function of time and subarray is then
\begin{equation}
C(t) = \frac{A + B \exp \left[ -(t - t_o)/\tau \right] } {D_\mathrm{SA}}
\end{equation}
where $t$ is the time in MJD, $A$ gives the calibration factor at infinite time after the exponential has fully decayed, $B$ is the amplitude of the time dependent term, $t_o = 59720~d$, $\tau$ is the time constant in days, and $D_\mathrm{SA}$ is the relative change with subarray.
Tables~\ref{tab:calfac} and \ref{tab:subtrans} give the coefficients for each band and subarray, respectively.
For the coronagraphic and FND filters, the temporal dependence was calculated by interpolating between the two imaging filters to either side in wavelength and weighting by the distance in wavelength.
The number of stars contributing to the calibration factors $n_\mathrm{stars}$ can be non-integral if some of the observations for a star are not used in the average.
The final column gives the percentage repeatability $\sigma(\mathrm{repeat})$ as measured in \S\ref{sec:repeat}.

\begin{figure*}[tbp]
\epsscale{1.2}
\plotone{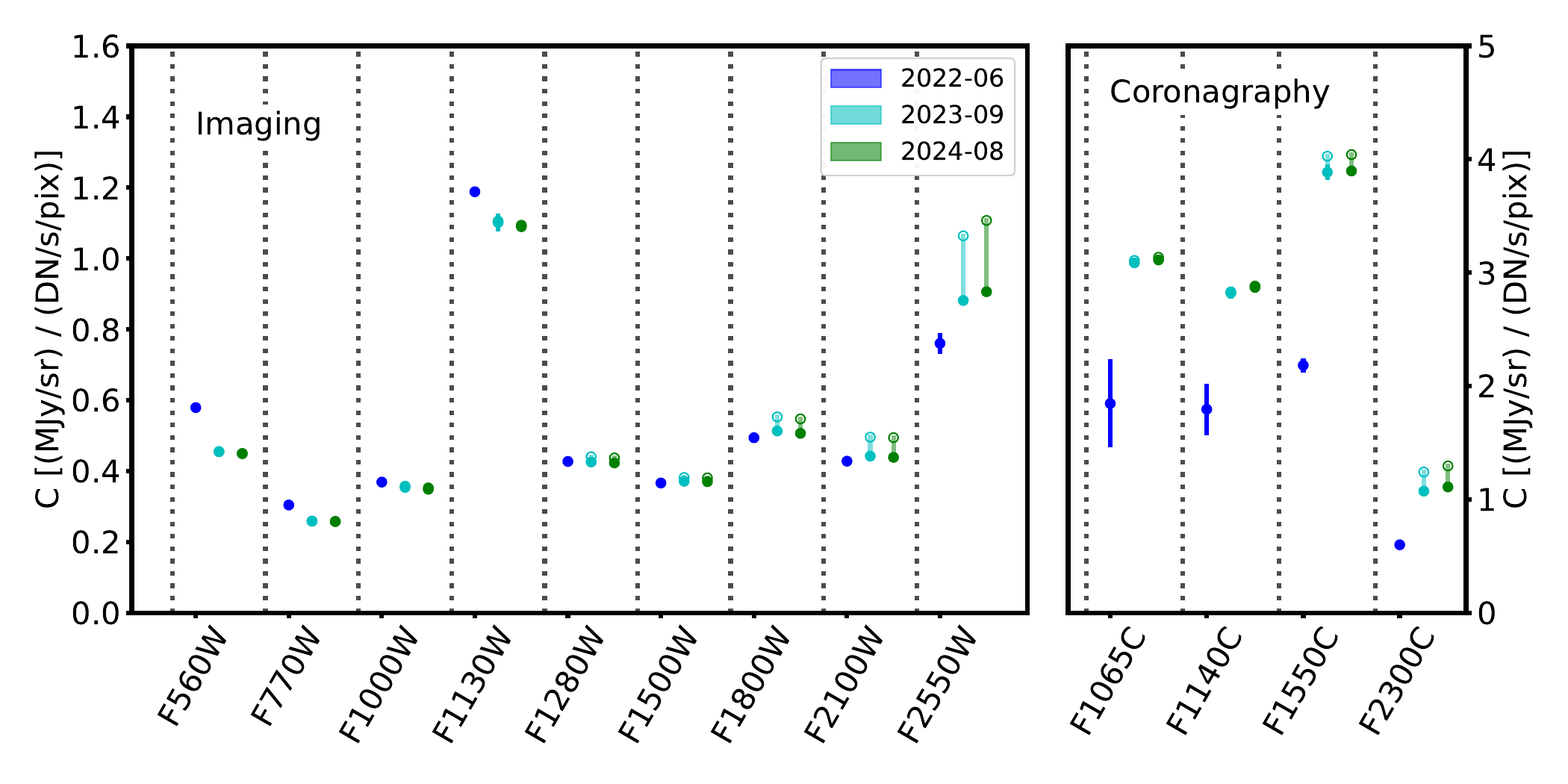}
\caption{The delivered calibration factors used by the pipeline are plotted versus time for all bands except FND.
The FND calibration has been determined for the first time with this work and so is not plotted.
For the two most recent deliveries, the time-dependent variation is shown as a solid symbol (start of observations) connected to an open symbol (asymptotic value after all expected change).
The changes reflect the improved calibration factor measurements with more observations and a better understanding of the telescope and instrument.
\label{fig:calfacs_delivered}}
\end{figure*}

Fig.~\ref{fig:calfacs_delivered} shows the refinement of the calibration factor measurements over time.
The first delivery in 2022 was based on Commissioning observations of 1--2 stars and the instrument knowledge at that time.
The second delivery in 2023 reflected observations of a significantly larger sample of stars and, critically, a better understanding of the instrument, especially how to model the PSF.
For this second delivery, the PSFs were modeled with a combination of observed PSFs for the cores and model PSFs for the wings (like \S\ref{sec:apcor}).
In particular, the changes seen in the F560W and F770W bands were due to the inclusion of a model for the cruciform artifact that significantly affects these two bands.
In addition, the large change for all the coronagraphic bands resulted from correcting an error in the normalization of the model PSFs.
The third delivery is based on the results reported in this work and includes more observations and the addition of the subarray correction.

The standard deviation of the calibration factor $\sigma(CF)$ ranges from 0.32 to 0.98\% for the imaging bands and 0.42 to 0.90\% for the coronagraphic bands.
The $\sigma(CF)$ expressed as a percentage gives the contribution that the absolute flux calibration makes to the uncertainty on any measurement expressed in physical units.
The repeatability uncertainty, $\sigma(\mathrm{repeat})$, gives the uncertainty for any single measurement taken with a similar four-point dither measurement of a point source.
In other words, these two uncertainties should be combined in quadrature with the uncertainties based on read, photon, and flatfield uncertainties reported by the JWST pipeline.

The uncertainties quoted are correct for measurements of a well-exposed point source, in other words observations like those taken for the flux-calibration stars.
In particular, the uncertainties for resolved sources that focus on surface brightnesses are larger, because the surface brightness uncertainties include the uncertainties in the aperture corrections to infinite aperture.
The uncertainties quoted in \S\ref{sec:apcor} for the aperture corrections are small and are based mainly on scatter between similar measurements.
These uncertainties do not probe the systematic uncertainties related to effects not present in the PSF models.
In particular, the modeling of the cruciform for the F560W and F770W bands is known to be incomplete, which contributes an additional uncertainty estimated to be on the order of 3\% based on an independent analysis using highly saturated PSFs.

\section{Summary \label{sec:summary}}

The absolute flux calibration for MIRI Imaging and Coronagraphy is based on over two years of dedicated observations.
These observations include stars of three different types: eight hot stars, eleven A dwarfs, and nine Solar analogs.

The main components of the calculated calibration factors are the predicted fluxes for each star, aperture photometry for each observation in instrumental units, and aperture corrections to provide the correction to infinite aperture.
Well-exposed, isolated stars observed as part of the absolute flux calibration were combined with model PSFs and used to measure the enclosed energy curves, determine the radii, sky annuli, and aperture corrections for aperture photometry.
Automated routines\footnote{\url{https://github.com/STScI-MIRI/ImagingFluxCal}} produced all of the aperture photometry, allowing for straightforward updating of the photometry with new JWST pipeline versions and calibration reference files.
The predicted fluxes for each star were determined by integrating the predicted fluxes from models with the appropriate band response functions.

Calibration factors for each observation were computed and examined for any dependencies.
Dedicated repeat measurements of one star supplemented by observations of two other stars revealed that the longer-wavelength imaging filters suffer from a time-dependent loss in total system response.
This loss is fit well with an exponential model with a time constant around 200~days with scatter less than 1\% for all but the F2550W band.
F1280W and longer-wavelength bands were seen to have a significant loss, starting at 3\% and rising to 22\% at F2550W.
After correcting for the temporal response loss, a subarray dependence was found by combining dedicated observations of two stars observed in all of the subarrays with all the average calibration factor measurements for each subarray.
The dependence relative to FULL frame ranges from no change to 3.4\% for the SUB64 subarray.

A search for dependencies with other source and detector parameters reveals no significant dependencies after correction for the known temporal and subarray dependencies.
Source dependencies checked were versus predicted model flux, measured background, and type of star.
In addition, the accuracy of the time-dependent correction was confirmed to work well for all the observations.
The detector dependencies were checked versus the central-pixel rate, central-pixel well depth, and the integration time.
In addition, the accuracy of the subarray correction was tested and found to work well.

The time- and subarray-dependent calibration factors for each imaging and coronagraphic band are reported and have been delivered for use in the JWST pipeline.
For the first time, this work provides the calibration factor for the wide neutral-density FND band that is used for target acquisition.
The scatter in the calibration factors from individual stars ranges from 1 to 4\% depending on the band.
The uncertainty in the average calibration factors ranges from 0.3 to 1.0\% depending on the band.
The uncertainties on the calibration factors are lower than pre-launch expectations (e.g., ``requirements'') and support a wide range of astrophysical investigations including those needing percent-level or better absolute-flux calibration.

The imaging and coronagraphic flux calibration will continue to improve as additional observations of absolute flux calibration targets are taken and the knowledge of the MIRI instrument improves.

\begin{acknowledgements}

This work is based [in part] on observations made with the NASA/ESA/CSA James Webb Space Telescope. The data were obtained from the Mikulski Archive for Space Telescopes at the Space Telescope Science Institute, which is operated by the Association of Universities for Research in Astronomy, Inc., under NASA contract NAS 5-03127 for JWST. These observations are associated with programs \#1027, 1045, 1523, 1524, 1536, 1537, 1538, 1539, 4488, 4496, 4497,
96 4498, 4499, 4578, and 6607.
The following National and International Funding Agencies funded and supported the MIRI development: NASA; ESA; Belgian Science Policy Office (BELSPO); Centre Nationale d'Etudes Spatiales (CNES); Danish National Space Centre; Deutsches Zentrum fur Luftund Raumfahrt (DLR); Enterprise Ireland; Ministerio De Economi´a y Competividad; Netherlands Research School for Astronomy (NOVA); Netherlands Organisation for Scientific Research (NWO); Science and Technology Facilities Council; Swiss Space Office; Swedish National Space Agency; and UK Space Agency.
MD acknowledges support from the Research Fellowship Program of the European Space Agency (ESA).

\software{Astropy \citep{astropy:2013, astropy:2018, astropy:2022}, Photutils \citep{Bradley22}}

\end{acknowledgements}

\appendix

\section{All Calibration Factor Plots}
\label{sec:calfac_plots}

Figs.~\ref{fig:calfacs_mflux}--{\ref{fig:calfacs_subarr} show the calibration factor plots for all the imaging and coronagraphic bands for all the source and detector parameters discussed in Sections~\ref{sec:srcdep} and \ref{sec:detector}.
These allow the behaviors discussed in the above referenced sections to be investigated for all bands as well as a straightforward comparison of behaviors between bands.

\begin{figure*}[tbp]
\epsscale{1.15}
\plotone{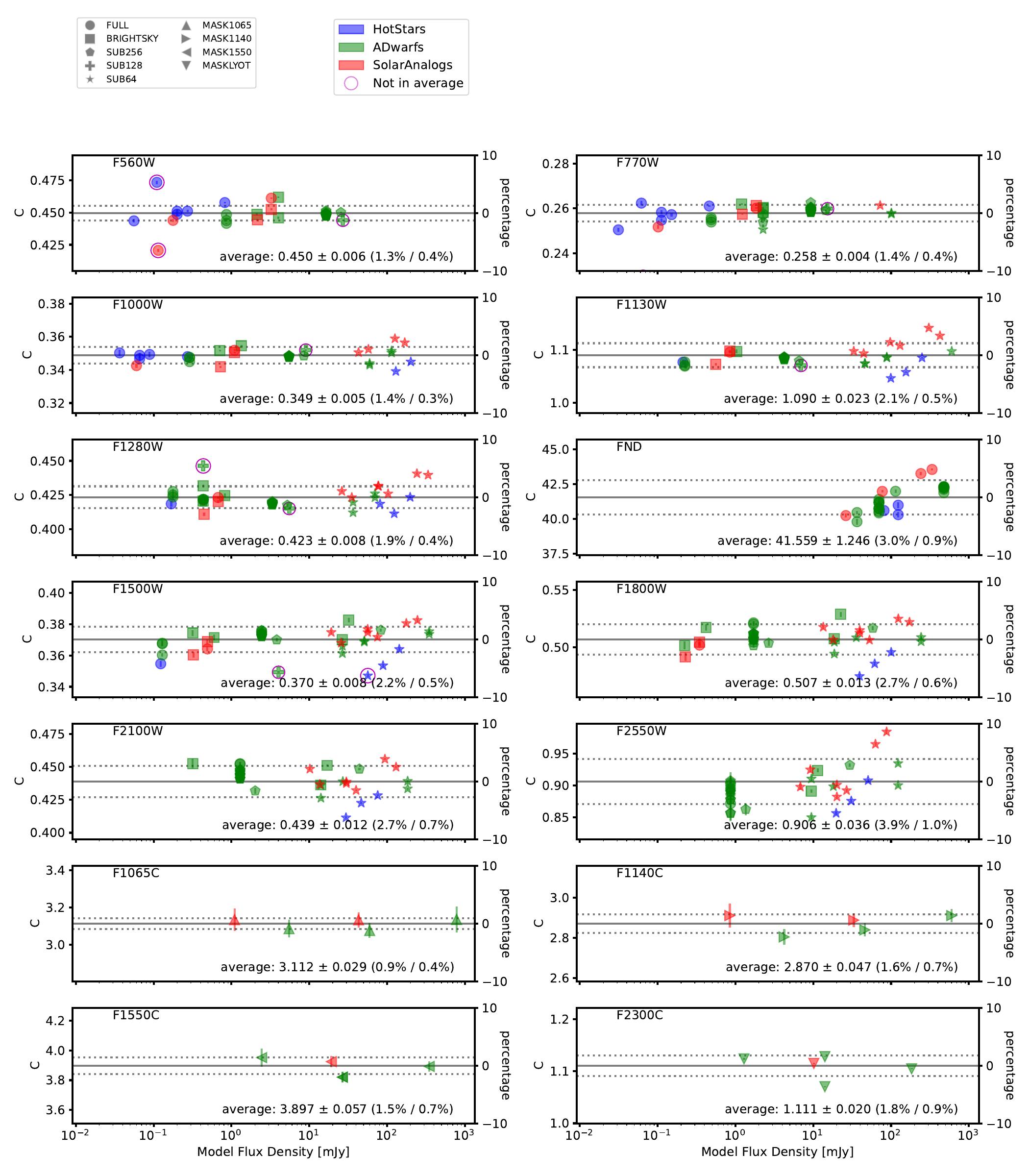}
\caption{The calibration factors as a function of model flux densities for every observation for the imaging and coronagraphic filters.
The temporal response loss and subarray variations have been corrected.
While the observations of HD~180609 are within the scatter for some bands, it is flagged in all bands and not used (\S\ref{sec:stars_removed}).
In addition, some observations are flagged as $> 3.5\sigma$ from the weighted average and not used (\S\ref{sec:stars_removed}). 
A horizontal solid gray line and the parallel dotted lines give the weighted average $\pm 1\sigma$ standard deviation.
The calibration factors and uncertainties are given in the lower right of each plot, and the percentage standard deviation and standard deviation of the mean are given in parentheses as well.
\label{fig:calfacs_mflux}}
\end{figure*}

\begin{figure*}[tbp]
\epsscale{1.15}
\plotone{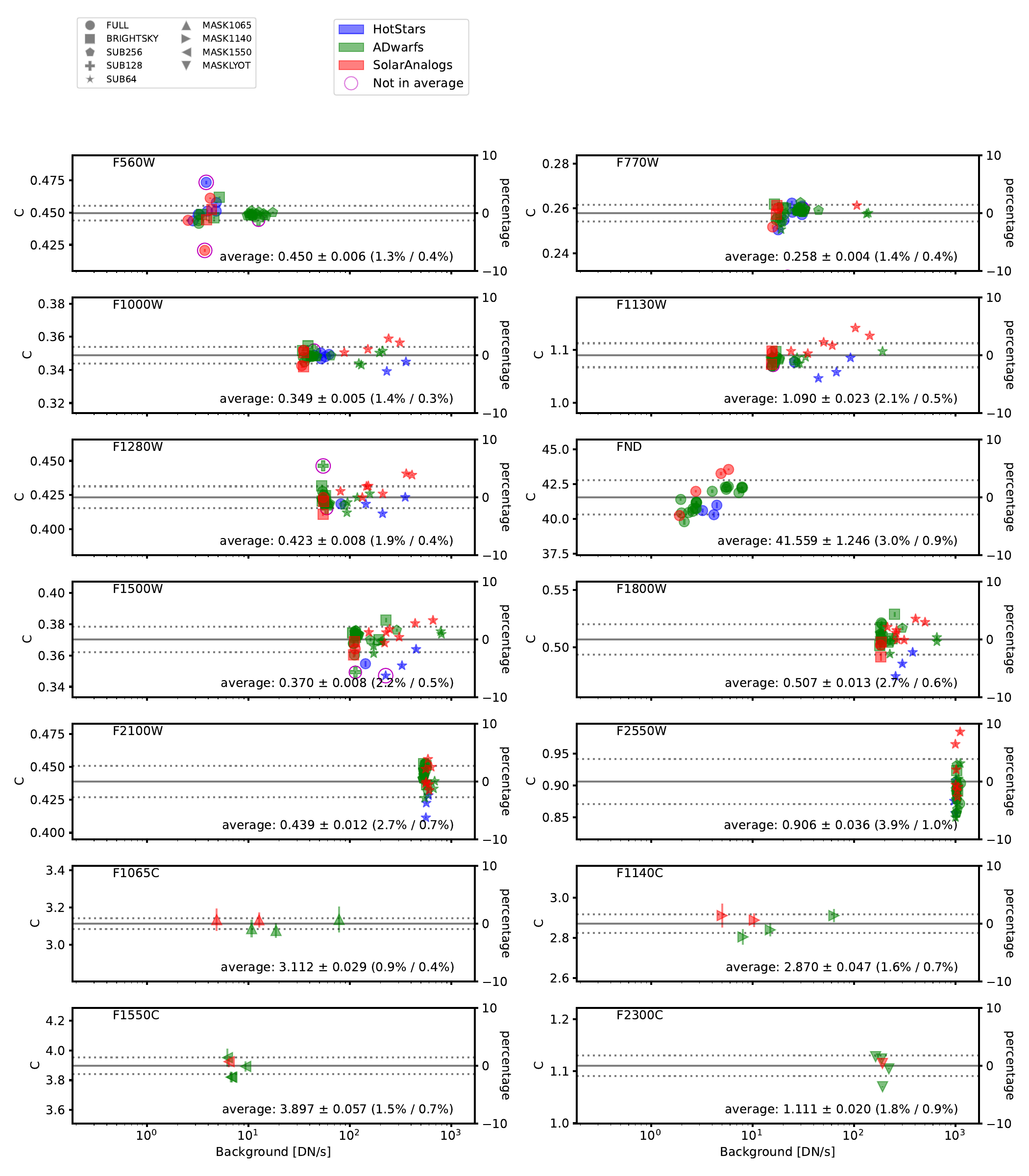}
\caption{Same as Fig.~\ref{fig:calfacs_mflux} except plotted versus measured background.
\label{fig:calfacs_bkg}}
\end{figure*}

\begin{figure*}[tbp]
\epsscale{1.15}
\plotone{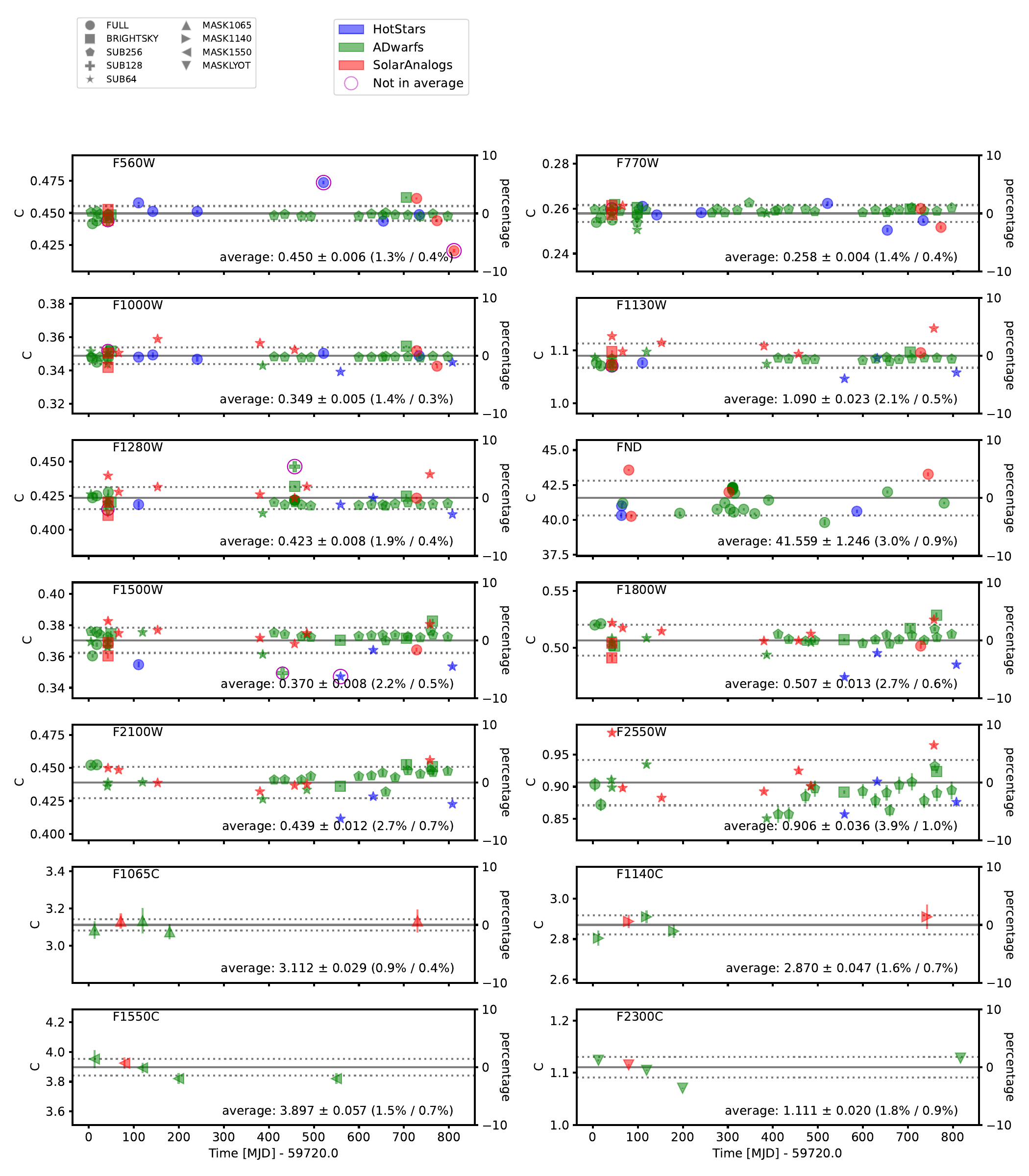}
\caption{Same as Fig.~\ref{fig:calfacs_mflux} except plotted versus observation time.
\label{fig:calfacs_time}}
\end{figure*}

\begin{figure*}[tbp]
\epsscale{1.15}
\plotone{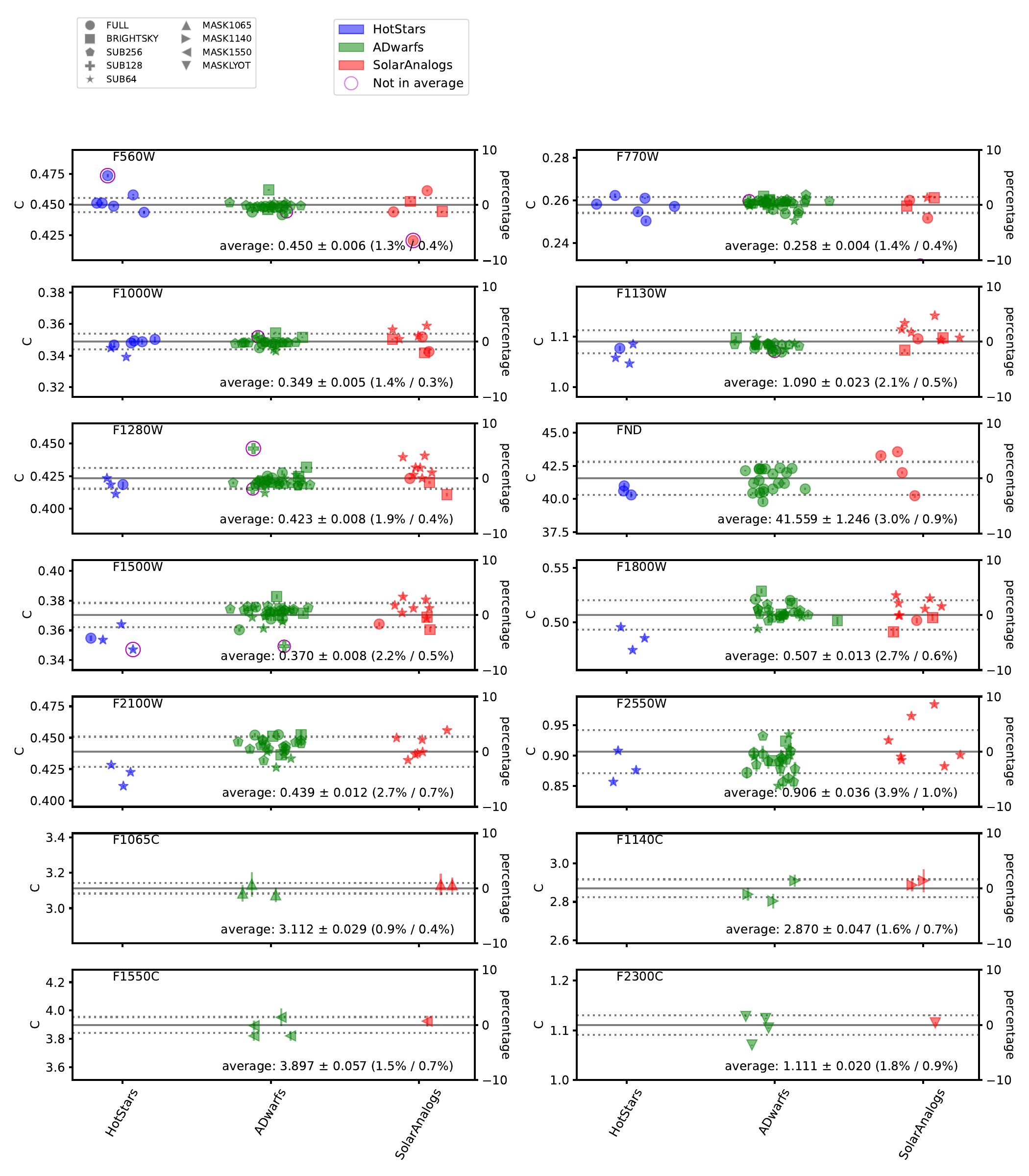}
\caption{Same as Fig.~\ref{fig:calfacs_mflux} except plotted versus source type.
\label{fig:calfacs_srctype}}
\end{figure*}

\begin{figure*}[tbp]
\epsscale{1.15}
\plotone{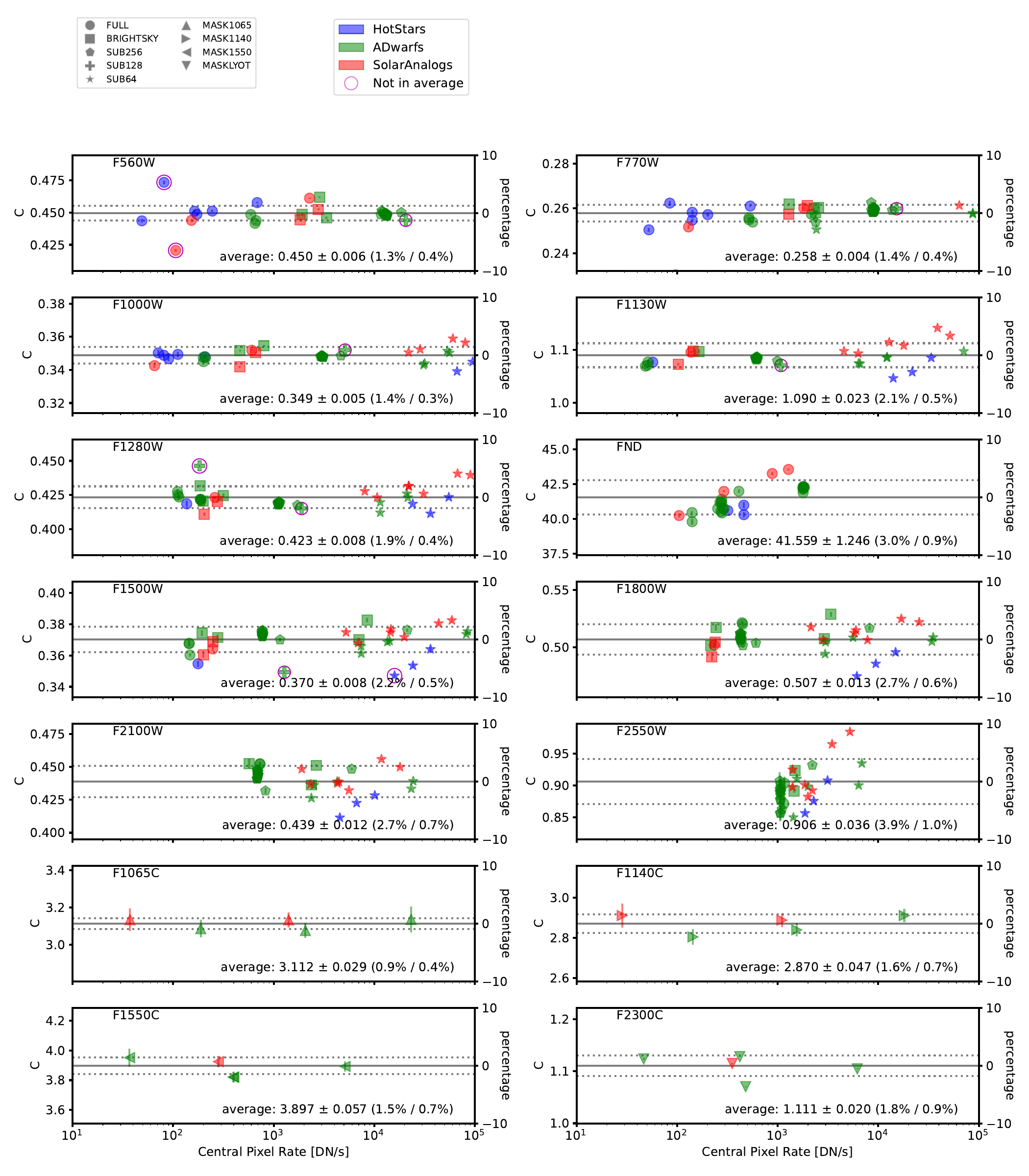}
\caption{Same as Fig.~\ref{fig:calfacs_mflux} except plotted versus central pixel rate.
\label{fig:calfacs_rate}}
\end{figure*}

\begin{figure*}[tbp]
\epsscale{1.15}
\plotone{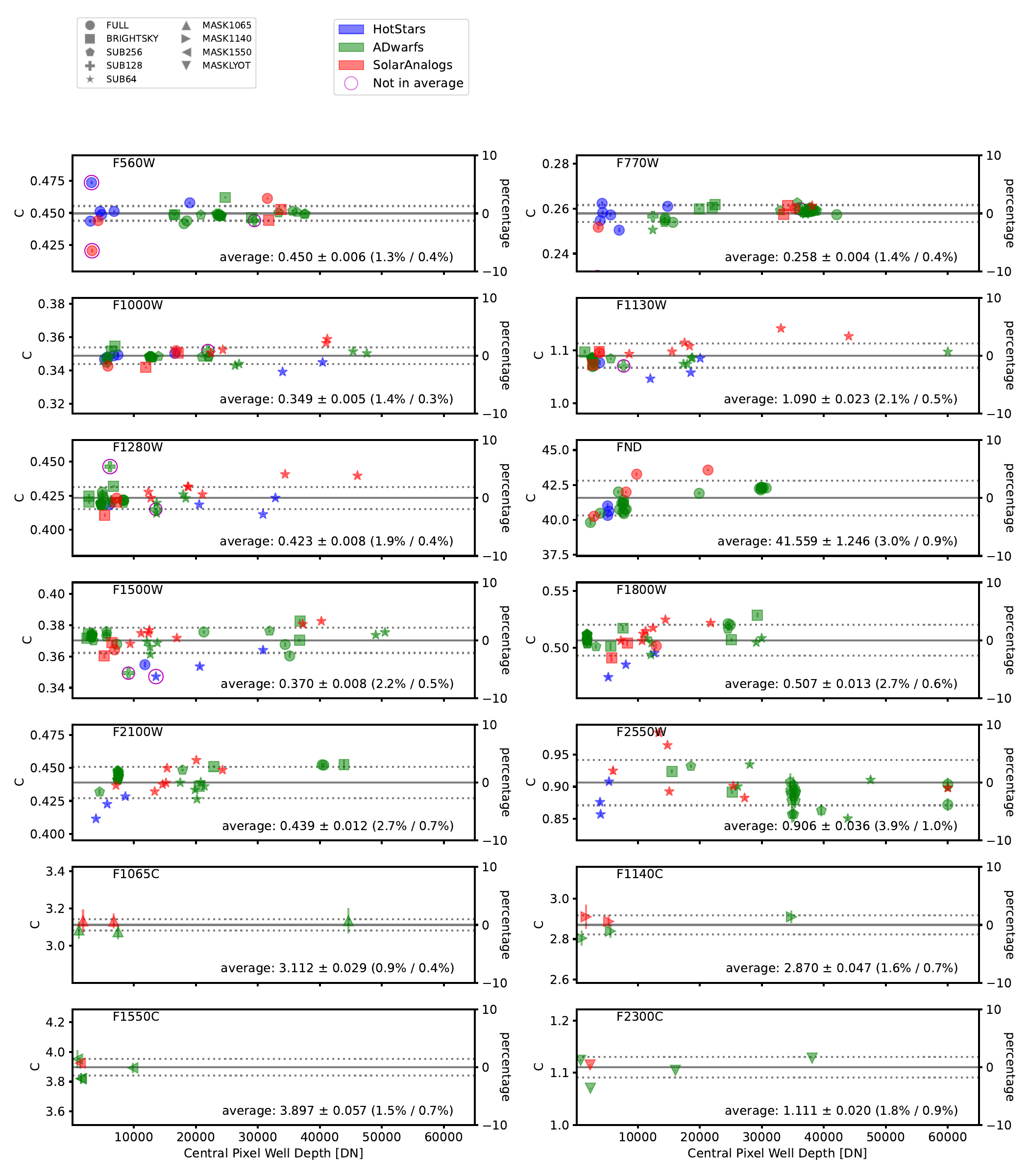}
\caption{Same as Fig.~\ref{fig:calfacs_mflux} except plotted versus central pixel well depth.
\label{fig:calfacs_welldepth}}
\end{figure*}

\begin{figure*}[tbp]
\epsscale{1.15}
\plotone{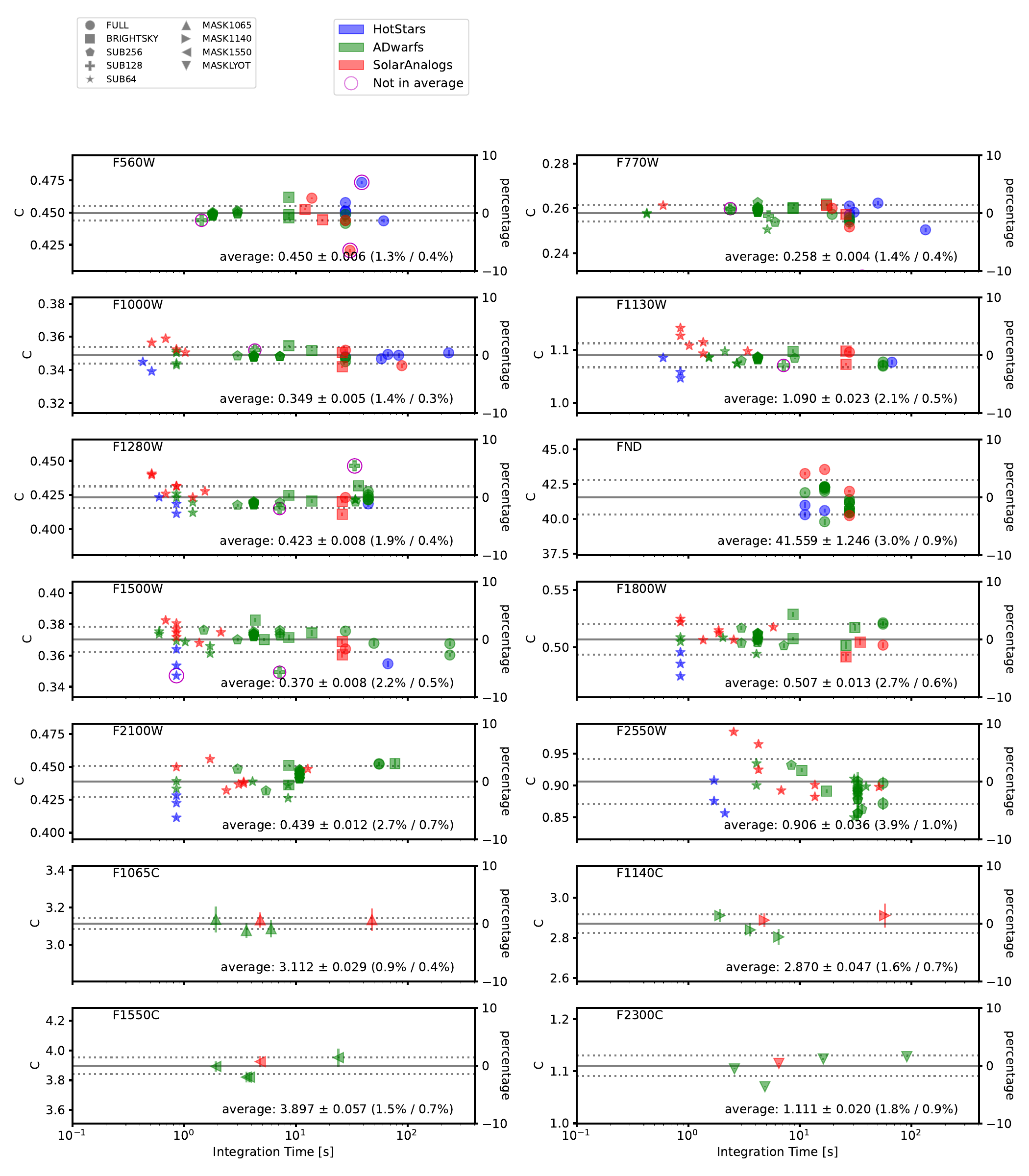}
\caption{Same as Fig.~\ref{fig:calfacs_mflux} except plotted versus integration time.
\label{fig:calfacs_inttime}}
\end{figure*}

\begin{figure*}[tbp]
\epsscale{1.15}
\plotone{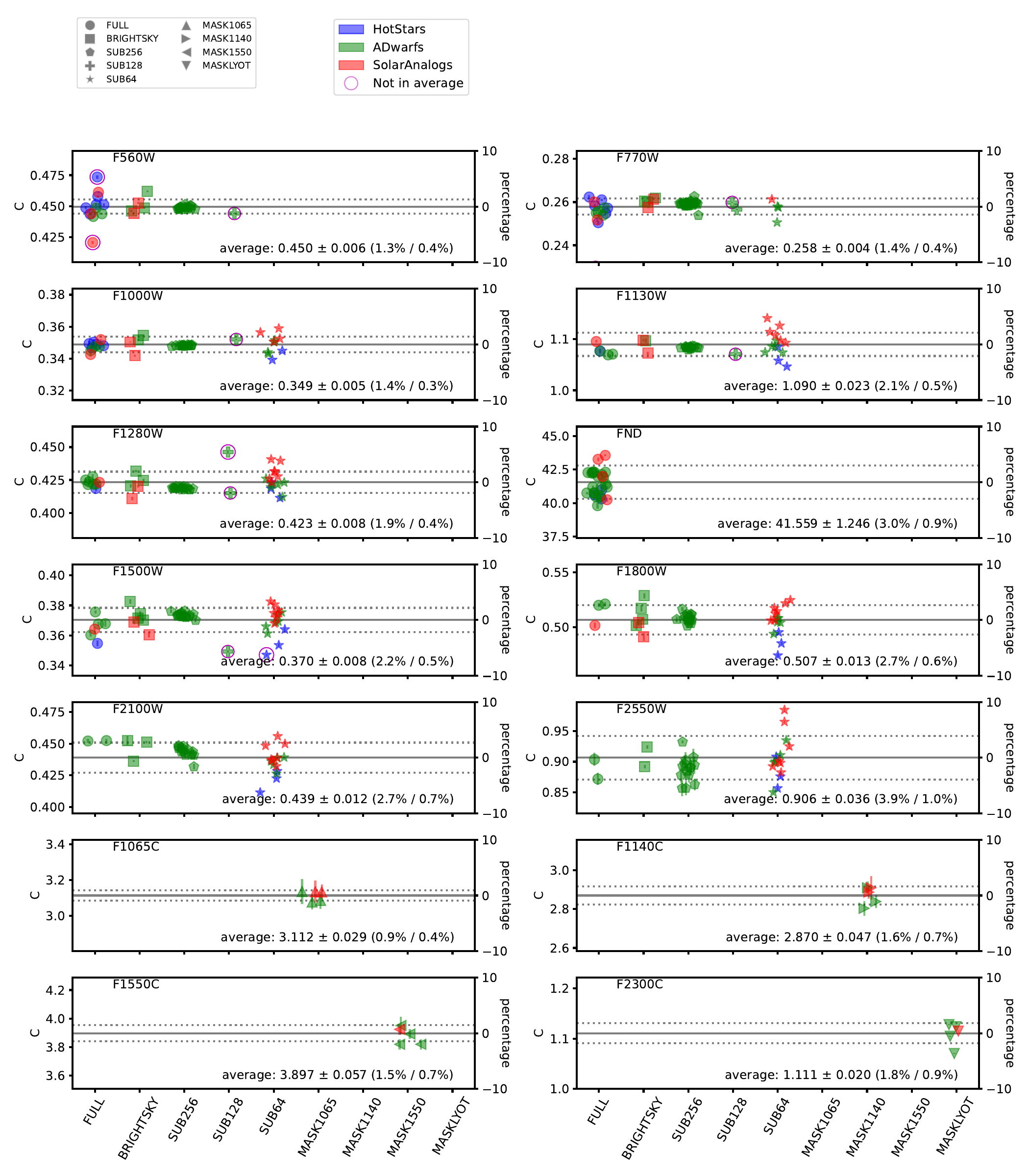}
\caption{Same as Fig.~\ref{fig:calfacs_mflux} except plotted versus subarray.
\label{fig:calfacs_subarr}}
\end{figure*}

\end{document}